\documentclass[%
aip,
jmp,
amsmath,amssymb,
reprint,%
]{revtex4-2}

\usepackage{graphicx}
\usepackage{dcolumn}
\usepackage{bm}

\usepackage[utf8]{inputenc}
\usepackage[T1]{fontenc}
\usepackage{mathptmx}
\usepackage{color}
\usepackage{etoolbox}
\usepackage{braket}
\graphicspath{{GraphenDiss}}
\usepackage{amsmath}
\usepackage{dsfont}
\usepackage{mathtools}
\usepackage{hyperref}
\usepackage{comment}
\usepackage{amsmath}
\usepackage{amssymb}
\usepackage[paper=a4paper,left=25mm,right=25mm,top=25mm,bottom=25mm]{geometry}
\usepackage{mathtools}
\usepackage{dsfont}
\usepackage{hyperref} 
\usepackage{cleveref}
\crefname{figure}{Fig.}{Fig.}
\crefname{equation}{Eq.}{Eq.}
\usepackage{siunitx}
\usepackage{subcaption}
\usepackage{graphicx} 
\usepackage[sort&compress]{natbib}
\graphicspath{{GraphenDiss}}
\usepackage{float}
\usepackage{orcidlink}

\makeatletter
\def\@email#1#2{%
	\endgroup
	\patchcmd{\titleblock@produce}
	{\frontmatter@RRAPformat}
	{\frontmatter@RRAPformat{\produce@RRAP{*#1\href{mailto:#2}{#2}}}\frontmatter@RRAPformat}
	{}{}
}%
\makeatother
\begin{document}
	
	\preprint{AIP/123-QED}
	
	\title[The Ericson Transition in Time--Reversal Invariant Systems]{The Ericson Transition in Time--Reversal Invariant Systems: Orthogonally Invariant Hamiltonians}
	
	\author{Simon Köhnes$^{1}$ and Thomas Guhr$^1$}
	\affiliation{
		$^1$ Fakult\"at f\"ur Physik, Universit\"at Duisburg--Essen, Duisburg, Germany
	}
	\email{simon.koehnes@uni-due.de}
	
	\date{\today}
\begin{abstract}
	We investigate the distributions of off-diagonal scattering matrix elements and cross-sections in the Ericson regime for systems that are time--reversal and rotationally invariant.
	In previous works, we developed techniques to analyze the Ericson transition by means of asymptotic expansions in the cases of $\beta=2,\ 4$. Here, we prove that Gaussian and exponential behavior is found as the leading--order term of an asymptotic expansion of those distributions also in the orthogonal case $\beta=1$, which is often the relevant one for analyzing experiments. Moreover, we derive the subleading--order term and compare with numerical simulations. Finally, we discuss similarities and differences of the Ericson transition for all three Wigner--Dyson classes.
\end{abstract}
\maketitle
	\section{Introduction}
	Scattering theory is a cornerstone of quantum physics, both from an experimental and a theoretical viewpoint. Many scatttering systems, including classical and quantum systems, exhibit chaotic or stochastic behavior, for numerous examples see \cite{LetterUniversality} and references therein. A typical sign of chaotic scattering is the Ericson regime, in which the resonances of a system with many open scattering channels strongly overlap. The cross sections then exhibit random behavior in the sense that they become exponentially distributed and the scattering matrix elements follow a universal Gaussian distribution. This was proven for the case of time--reversal noninvariant system in \cite{LetterUniversality} and recently  for the case of time--reversal invariant systems with half--integer spins.
	
	We now proceed as in the work of \cite{LetterUniversality} and derive the universal Gaussian for time--reversal and rotationally invariant systems. We also calculate subleading--order corrections to the Gaussian. These characterize the transition region where $\Xi \simeq 1$. We also give explicit analytical results for the cross--section distribution and the moments. 
	We infer the methodology and notation as presented in \cite{LetterUniversality,Matrixelemente,Wirkungsquerschnitte}, such that the Hamilton operator $H$ is chosen as a random matrix from the Gaussian Orthogonal Ensemble GOE $(\beta=1)$ with Gaussian probability density
	\begin{align}
		P_\beta(H)\sim \exp{\Biggl(-\frac{\beta N}{4\nu^2}\text{tr} H^2\Biggr)}.
	\end{align}
	The Ericson transition is then characterized by the ratio of average resonance
	width $\Gamma$ and average mean level spacing $D$, \textit{i.e.} by the
	dimensionless parameter
	\begin{align}
		\Xi= \frac{\Gamma}{D}= \frac{1}{2\pi} \sum\limits_{c=1}^{M}T_c.
	\end{align} We extend the results from \cite{LetterUniversality}, to solve the Ericson transition by means of asymptotic
	expansions in powers of $1/ \Xi$. Here, we present the technically most challenging case of $\beta=1$. We proceed as follows: We derive the Gaussian distribution of real-- and imaginary parts of $S$--matrix elements in Ericson regime. Upon doing so, we derive asymptotic corrections to the universal Gaussian as in the cases of $\beta=2,\ 4$. Finally, we draw a comparison between these cases.
	\section{Ericson transition in the case of $\beta=1$}
In Sec. II.A we formulate the problem and give the exact result for the characteristic function $R_s(k)$. In Sec. II.B, we consider the manifold structure and do a channel--factor approximation. Thereafter, in Sec. II.C, we derive asymptotes for Bessel functions via a multinomial expansion. Finally, in Sec. II.D, we derive the universal Gaussian for $\beta=1$.
	\subsection{Characteristic function}
	We want to find asymptotic expansions in powers of $1/\Xi$ for the distribution $P_s(x_s)$ of real-- and imaginary parts of $S$--matrix elements, \textit{e.g.} $x_1=\text{Re}(S_{ab}),\ x_2=\text{Im}(S_{ab})$.
	Starting point of our analysis is, as in our previous works,  the univariate characteristic function, which also serves as moment generating function, \textit{e.g.},  
	\begin{align}
		R_s(k) = \int\limits_{-\infty}^{\infty}dx_s P_s(x_s)\exp{(-ikx_s)}=\sum\limits_{n=0}^{\infty}\frac{(-ik)^n}{n!}\overline{x_s^n}.
	\end{align}
	The exact analytical expression for the univariate characteristic functions for $\beta=1$ was derived in \cite{Matrixelemente} and is given by
	\begin{align}\label{eq:Char}
		R_s(k) = 1 +\frac{1}{8 \pi} \int\limits_{-1}^{1}d \lambda_0 \int\limits_{1}^{\infty}d\lambda_1 \int\limits_{1}^{\infty}d\lambda_2 \int\limits_{0}^{2\pi} d\psi J(\lambda_0,\lambda_1 , \lambda_2) \mathcal{F}_O(\lambda_0, \lambda_1,\lambda_2)(\kappa_1 k +\kappa_2 k^2 +\kappa_3 k^3 +\kappa_4 k^4).
	\end{align}
		In the characteristic function $R_s(k)$, the coefficients of the Bessel functions are given in terms of the functions
	\begin{align}\label{eq:6.42}
		p_c^0 &= \frac{\sqrt{1-\lambda_0^2}}{8(g_c^+ + \lambda_0)} ,\\
		p_c^1&=\frac{\sqrt{\lambda_1^2-1}}{8(g_c^+ + \lambda_1)},\\
		p_c^2&=\frac{\sqrt{\lambda_2^2-1}}{8(g_c^+ + \lambda_2)},\\
		p_c^{\pm} &= p_c^1\pm p_c^2.
	\end{align}
	
	Furthermore, we have the quantities
	\begin{align}\label{eq:6.43}
		q_c^+ &= \frac{(-i)^{s-1}}{8}\left(\frac{E}{\Delta}+i g_c^{-}\right)\left(\frac{1}{g_c^+ +\lambda_1} +\frac{1}{g_c^+ +\lambda_2}-\frac{2}{g_c^+ +\lambda_0}   \right),
	\end{align}
	and
	\begin{align}\label{eq:6.44}
		q_c^- &= \frac{(-i)^{s-1}}{8}\left(\frac{E}{\Delta}+i g_c^{-}\right)\left(\frac{1}{(g_c^+ +\lambda_1)}-\frac{1}{(g_c^+ +\lambda_2)}\right).
	\end{align}
	We also have
	\begin{align}
		X \label{eq:11}&= 2p_a^+ +q_a^- \exp{(-2\psi i)} + r_a^- \exp{(2\psi i)},\ Y= 2p_b^+-q_b^-\exp{(2\psi i) }-r_b^-\exp{(-2\psi i)},\\
		\omega &= 2\sqrt{XY},\\
		l &= \frac{X}{Y} = \frac{2p_a^+ +q_a^- \exp{(-2\psi i)} + r_a^- \exp{(2\psi i)}}{2p_b^+-q_b^-\exp{(2\psi i) }-r_b^-\exp{(-2\psi i)}},\\
		m &= \frac{Y}{X}=\frac{2p_b^+-q_b^-\exp{(2\psi i) }-r_b^-\exp{(-2\psi i)}}{2p_a^+ +q_a^- \exp{(-2\psi i)} + r_a^- \exp{(2\psi i)}},
	\end{align}
	and $r_c^{\pm}=(q_c^\pm)^*$, where the star denotes complex conjugation. The $\kappa_j$ are defined by
	
	\begin{align}
		\kappa_{1} &= \kappa_{11}J_1(k\omega),\ &&\kappa_2 = \kappa_{21}J_0(k\omega)+\kappa_{22}J_2(k\omega),\\
		\kappa_3 &=\kappa_{31}J_1(k\omega)+\kappa_{32}J_3(k\omega),\ &&\kappa_4 = \kappa_{41}J_0(k\omega)+\kappa_{42}J_2(k\omega) +\kappa_{43}J_4(k\omega).
	\end{align}
	The explicit form of the coefficients $\kappa_{ij}$ is given in \textit{Appendix A}. We now proceed completely analogous to the aforementioned cases of $\beta=2,\ 4$ and derive leading-- and subleading--order asymptotes of the moments in $1/\Xi$. By resummation of the moments, we reconstruct the asymptotes of the characteristic function, which, upon inverse Fourier transform and rescaling, yields the asymptotic expansion in $1/\Xi$ for the distributions $P_s(\xi_s)$, where $\xi_s = x_s/\sqrt{\Xi}$. Yet, the technical problems to counter are much more challenging than in the previously discussed cases and will be considered in detail. 
	We have to deal with four problems. First, the Berezinian 
	\begin{align}
		J(\lambda_0,\lambda_1,\lambda_2) = \frac{(1-\lambda_0^2)|\lambda_1-\lambda_2|}{2(\lambda_1^2-1)^{1/2}(\lambda_2^2-1)^{1/2}(\lambda_1-\lambda_0)^2(\lambda_2-\lambda_0)^2}
	\end{align}
	shows a coordinate singularity at $\lambda_0=\lambda_1=\lambda_2=1$. This behavior is, at least in the basis of singular values $\lambda_j$, only partially countered by zeros of the remaining integrand. However, this issue amended by choosing different coordinates similar to the symplectic case. Second, to generate the asymptotic expansion, the channel factor 
	\begin{align}
		\mathcal{F}_0(\lambda_0,\lambda_1,\lambda_2)&=\prod\limits_{c=1}^{M}\frac{g_c^++\lambda_0}{(g_c^++\lambda_1)^{1/2}(g_c^++\lambda_2)^{1/2}}
	\end{align}
	must be written as an exponential,
	\textit{e.g.}
	\begin{align}
	\mathcal{F}_0(\lambda_0,\lambda_1,\lambda_2)\simeq \exp{\left(-\frac{\pi\Xi}{2}(\lambda_1+\lambda_2-2\lambda_0)\right)}
	\end{align}
	as shown in \textit{Appendix B}. This turns the characteristic function into a Laplace--type integral.
	Third, in contrast to the cases $\beta=2,\ 4$ we have to deal with an additional integral over the phase $\psi$, which is a remnant of a former group integral. This problem is tackled simultaneously with the generation of the asymptotic expansion. Fourth and finally, we have to solve the remaining two--dimensional integrals. In the following, we go through these four steps in the orthogonal case. 
	\subsection{Manifold structure and channel--factor approximation}
	We start by eliminating the coordinate singularity by using the transformation
	\begin{align}\label{eq:Trafo}
		\lambda_0 &= 1- p's'^2(1-t'^2),\\
		\lambda_1 &=\label{eq:Trafo1} 1+p'(1+s'^2t'^2-2s'^2),\\
		\lambda_2 &=\label{eq:Trafo3}1+p's'^2t'^2  ,
	\end{align}
	which maps the singularity to $p'=0$. This is, up to linear transformations, the same transformation as used in \cite{VerbaarschotCorr}, where the average of two $S$--matrix elements in the $\beta=1$ case was investigated. The phase $\psi$ is not contributing to the singularity structure and must therefore not be reparameterized. The Jacobian of the transformation evaluates to 
	\begin{align}\label{eq:Trafo2}
		\left|\frac{\partial (\lambda_0,\lambda_1,\lambda_2)}{\partial(s',t',p')}\right|&= 4 p'^2 s'^3t'.
	\end{align}
	The Jacobian combined with contributions of the remaining integrand lifts the coordinate singularity at $p'=0$. 
	Inversion of the coordinate transformation yields 
	\begin{align}
		p' &= \lambda_1+\lambda_2-2\lambda_0, \\\nonumber
		s' &= \sqrt{\frac{\lambda_2 - \lambda_0}{\lambda_1 + \lambda_2 -2\lambda_0}},\\\nonumber
		t' &= \sqrt{\frac{\lambda_2-1}{\lambda_2 - \lambda_0}}
	\end{align}
	The integration domain is now determined by the inequalities
	\begin{alignat}{3}
		0&\leq  p's'^2(1-t'^2)&&\leq2\label[ineq]{ineq:I}\\
		0&\leq   p's'^2t'^2&&< \infty \label[ineq]{ineq:II}\\
		0&\leq  p's'^2t'^2 &&\leq p'(1+s'^2t'^2-2s'^2)\label[ineq]{ineq:III}.
	\end{alignat} For $0\leq p' \leq4$ we have 
	\begin{align}
		0\leq p'\leq 4,\ 0\leq s'\leq \frac{1}{\sqrt{2}},\ 0\leq t' \leq 1,
	\end{align}
	thus, the integration over each variable in this parametrization are done independently from one another. We refer to this as the disconnected part of the saddle--point manifold.
	For $p'>4$, the integration domain shows a more intricate geometric structure, to which we refer as the connected part of the saddle--point manifold. To generate an asymptotic expansion, however, it is sufficient to consider the disconnected part only, since for $p'>4$ the integrand in \cref{eq:Char} is decaying exponentially in $\Xi$ and is therefore negligible in the Ericson regime. This is similar to the cases of $\beta=2, 4$.
	By approximating the channel--factor as 
	\begin{align}\label{eq:exponential}
		\mathcal{F}_O(p',s',t') &\simeq \text{exp}\left(-\frac{\pi\Xi}{2}p'\right),
	\end{align}
	the characteristic function turns into a multidimensional Laplace--type integral. Thus, after generating the asymptotic expansion, we have to carry out three further integrations. It turns out to be advantageous to first calculate the integrals over $\psi$, before doing the calculations of $s',t'$. 
	\subsection{Multinomial expansion of Bessel functions} 
	Since we want to use Watson's lemma, we have to find the leading-- and subleading--order terms around $p'=0$ arising from the Bessel functions in the integrand. By virtue of the recurrence relation of the Bessel functions, \textit{e.g.} 
	\begin{align}
		J_{\nu}(x) = \frac{x}{2\nu}\left(J_{\nu+1}(x)+J_{\nu-1}(x)\right),
	\end{align} it is sufficient to consider even Bessel functions only.
	In the following, we derive asymptotes for $p'\ll 1$ and even-order Bessel functions  of the form 
	\begin{align}
		J_{\nu}\left(2 k \left(\sum\limits_{j=1}^{\infty}\alpha_{j}(s',t',\psi)p'^{(j+1)/2}\right)^{1/2}\right).
	\end{align}
	This is a slightly different situation compared to the $\beta=4$ case, since the coefficient functions $\alpha_j$ now have an additional $\psi$--dependence and the powers of $p'$ are given by rational numbers, such that we have to do a multinomial expansion in contrast to the binomial expansion in the symplectic case. Furthermore, we now have to distinguish between the characteristic functions $R_1(k)$ and $R_2(k)$, because these differ under general circumstances.
	Again, we use the series representation of the Bessel functions and find
	\begin{align}
		J_{\nu}\left(2  k \left(\sum\limits_{j=1}^{\infty}\alpha_{j}(s',t',\psi)p'^{(j+1)/2}\right)^{1/2}\right) = \sum\limits_{n=0}^{\infty}\frac{(-1)^nk^{2n+\nu}}{n!(n+\nu)! }\left(\sum\limits_{j=1}^{\infty}\alpha_{j}(s',t',\psi)p'^{(j+1)/2)}\right)^{n+\nu/2}.
	\end{align}
	Since we are only interested in the leading-- an subleading-- order terms in $p'$ for $p'\ll 1$, we approximate by 
	\begin{align}
		\sum\limits_{j=1}^{\infty}\alpha_{j}(s',t',\psi)p'^{(j+1)/2}\simeq \alpha_{1}(s',t',\psi)p'+\alpha_{2}(s',t',\psi)p'^{3/2}+\alpha_{3}(s',t',\psi)p'^{2},
	\end{align}
	resulting in 
	\begin{align}
		\nonumber&\sum\limits_{n=0}^{\infty}\frac{(-1)^nk^{2n+\nu}}{n!(n+\nu)! }\left(\sum\limits_{j=1}^{\infty}\alpha_{j}(s',t',\psi)p'^{(j+1)/2)}\right)^{n+\nu/2}\\&\simeq 	\sum\limits_{n=0}^{\infty}\frac{(-1)^nk^{2n+\nu}}{n!(n+\nu)! }(\alpha_{1}(s',t',\psi)p'+\alpha_{2}(s',t',\psi)p'^{3/2}+\alpha_{3}(s',t',\psi)p'^{2})^{n+\nu/2}.
	\end{align}
	The term 
	\begin{align}
		(\alpha_{1}(s',t',\psi)p'+\alpha_{2}(s',t',\psi)p'^{3/2}+\alpha_{3}(s',t',\psi)p'^{2})^{n+\nu/2}
	\end{align}
	is now further evaluated by a multinomial expansion
	\begin{align}
		&\nonumber(\alpha_{1}(s',t',\psi)p'+\alpha_{2}(s',t',\psi)p'^{3/2}+\alpha_{3}(s',t',\psi)p'^{2})^{n+\nu/2}\\&\nonumber\simeq
		(\alpha_{1}(s',t',\psi)p')^{n+\nu/2}+\left(n+\frac{\nu}{2}\right)\alpha_{1}(s',t',\psi)^{n-1+\nu/2}\alpha_{2}(s',t',\psi)p'^{(2n+\nu+1)/2}\\&\nonumber+\Biggl(\frac{1}{2}\left(n+\frac{\nu}{2}\right)\left(n+\frac{\nu}{2}-1\right)\alpha_1(s',t',\psi)^{n+\nu/2-2}\alpha_2(s',t',\psi)^2+\left(n+\frac{\nu}{2}\right)\alpha_1(s',t',\psi)^{n+\nu/2-1}\alpha_3(s',t',\psi)\Biggr)\\&\times p'^{n+\nu/2+1}.
	\end{align}
	Thus, we have 
	\begin{align}
		\nonumber&\sum\limits_{n=0}^{\infty}\frac{(-1)^nk^{2n+\nu}}{n!(n+\nu)! }(\alpha_{1}(s',t',\psi)p'+\alpha_{2}(s',t',\psi)p'^{3/2}+\alpha_{3}(s',t',\psi)p'^{2})^{n+\nu/2}\\\nonumber&\simeq 		\sum\limits_{n=0}^{\infty}\frac{(-1)^nk^{2n+\nu}}{n!(n+\nu)! }\\\nonumber&\times\Biggl(	(\alpha_{1}(s',t',\psi)p')^{n+\nu/2}+\left(n+\frac{\nu}{2}\right)\alpha_{1}(s',t',\psi)^{n-1+\nu/2}\alpha_{2}(s',t',\psi)p'^{(2n+\nu+1)/2}\\\nonumber&+\Biggl(\frac{1}{2}\left(n+\frac{\nu}{2}\right)\left(n+\frac{\nu}{2}-1\right)\alpha_1(s',t',\psi)^{n+\nu/2-2}\alpha_2(s',t',\psi)^2+\left(n+\frac{\nu}{2}\right)\alpha_1(s',t',\psi)^{n+\nu/2-1}\alpha_3(s',t',\psi)\Biggr)\\&\times p'^{n+\nu/2+1}\Biggr).
	\end{align}
	Splitting the above sum into three, we calculate each sum individually. 
	First, we have 
	\begin{align}
		\sum\limits_{n=0}^{\infty}\frac{(-1)^nk^{2n+\nu}}{n!(n+\nu)! }(\alpha_{1}(s',t',\psi)p')^{n+\nu/2}= J_\nu(2k \sqrt{\alpha_{1}(s',t',\psi)p'}) = J_\nu^{(l)}(k\omega). 
	\end{align}
	For the second sum, we arrive at
	\begin{align}
		&\nonumber\sum\limits_{n=0}^{\infty}\frac{(-1)^nk^{2n+\nu}}{n!(n+\nu)! }\left(n+\frac{\nu}{2}\right)\alpha_{1}(s',t',\psi)^{n-1+\nu/2}\alpha_{2}(s',t',\psi)p'^{(2n+\nu+1)/2}\\&=\frac{\alpha_2(s',t',\psi) \sqrt{p'}}{2\alpha_1(s',t',\psi)}(\nu J_v(2k \sqrt{\alpha_1(s',t',\psi)p'})-2k\sqrt{\alpha_1(s',t',\psi)p'}J_{\nu+1}(2k\sqrt{\alpha_1(s',t',\psi)p'})) = J_\nu^{(sl)}(k\omega). 
	\end{align}
	For the third sum, we get 
	\begin{align}
		&\nonumber\sum\limits_{n=0}^{\infty}\frac{(-1)^nk^{2n+\nu}}{n!(n+\nu)! }\\&\nonumber\times\Biggl(\frac{1}{2}\left(n+\frac{\nu}{2}\right)\left(n+\frac{\nu}{2}-1\right)\alpha_1(s',t',\psi)^{n+\nu/2-2}\alpha_2(s',t',\psi)^2+\left(n+\frac{\nu}{2}\right)\alpha_1(s',t',\psi)^{n+\nu/2-1}\alpha_3(s',t',\psi)\Biggr)\\&\nonumber\times p'^{n+\nu/2+1}\\&\nonumber=\frac{p'}{8\alpha_{1}(s',t',\psi)}((4\alpha_1(s',t',\psi)\alpha_3(s',t',\psi)+\alpha_2(s',t',\psi)^2(\nu-2))\nu J_\nu(2k\sqrt{\alpha_1(s',t',\psi)p'})\\&\nonumber-4k(2\alpha_1(s',t',\psi)^{3/2}\alpha_3(s',t',\psi)\sqrt{p'}+\nu\alpha_2(s',t',\psi)^2\sqrt{\alpha_1(s',t',\psi)p'})J_{\nu+1}(2k\sqrt{\alpha_1(s',t',\psi)p'})\\&+4k^2p' \alpha_1(s',t',\psi)\alpha_2(s',t',\psi)^2J_{\nu+2}(2k\sqrt{\alpha_1(s',t',\psi)p'}))= J_\nu^{(ssl)}(k\omega).
	\end{align}
	Using these general expansions, the functions $\alpha_j(s',t',\psi)$ are inferred from the first three terms of a Taylor expansion of $ X Y$. First, for $R_1(k)$ we find, using \cref{eq:11}, that
	\begin{align}
		X Y &\nonumber= \frac{p'(s't'+\sqrt{1+s'^2t'^2-2s'^2})^2}{8(g_a^++1)(g_b^+1)}+p'^{3/2}(1-2s'^2)(s't'+\sqrt{1+s'^2t'^2-2s'^2})\\&\nonumber\times(E(\exp{(-2\psi i )}+\exp{(2\psi i)})(g_a^+-g_b^+) +i(\exp{(2\psi i)}-\exp{(-2\psi i)})(g_a^-(g_b^++1)+g_b^-(g_a^++1))\Delta)\\&\nonumber\times \frac{1}{16 \sqrt{2}(g_a^++1)^2(g_b^++1)^2\Delta}\\
		&\nonumber+p'^2\Biggl(\frac{g_a^+g_b^+-g_a^+-g_b^+-3}{16(g_a^++1)^2(g_b^++1)^2}\Biggl(1+2s'^2(t'^2-2)+2s'^4(2-2t'^2+t'^4)\\\nonumber&+s't'\sqrt{1+s'^2(t'^2-2)}+2s'^3t'(t'^2-1)\sqrt{1+s'^2(t'^2-2)}\Biggr)\\\nonumber&-\frac{E^2(1-2s'^2)^2\cos^2{(2\psi)}}{16(g_a^++1)^2(g_b^++1)^2\Delta^2}+\frac{g_a^-g_b^-(1-2s'^2)^2\sin^2{(2\psi)}}{16(g_a^++1)^2(g_b^++1)^2}\\&+\frac{E(1-2s'^2)^2\sin{(4\psi)}}{32(g_a^++1)^2(g_b^++1)^2\Delta}(g_b^--g_a^-)\Biggr)+\mathcal{O}(p'^{5/2}).
	\end{align}
	This expansion is now used with the formerly introduced multinomial expansion of the Bessel functions such that for $R_1(k)$ we find
	\begin{align}
		\alpha_1(s',t',\psi) &= \frac{(s't'+\sqrt{1+s'^2t'^2-2s'^2})^2}{8(g_a^++1)(g_b^++1)},\\
		\alpha_2(s',t',\psi)&\nonumber=((1-2s'^2)(s't'+\sqrt{1+s'^2t'^2-2s'^2}) )((\exp{(2\psi i)}+\exp{(-2\psi i)})(g_a^+-g_b^+) \\&\nonumber+i(\exp{(2\psi i)}-\exp{(-2\psi i)})(g_a^-(g_b^++1)+g_b^-(g_a^++1))\Delta)\\&\times \frac{1}{16 \sqrt{2}(g_a^++1)^2(g_b^++1)^2\Delta},\\
		\alpha_3(s',t',\psi)&\nonumber= \frac{g_a^+g_b^+-g_a^+-g_b^+-3}{16(g_a^++1)^2(g_b^++1)^2}\Biggl(1+2s'^2(t'^2-2)+2s'^4(2-2t'^2+t'^4)\\\nonumber&+s't'\sqrt{1+s'^2(t'^2-2)}+2s'^3t'(t'^2-1)\sqrt{1+s'^2(t'^2-2)}\Biggr)\\\nonumber&-\frac{E^2(1-2s'^2)^2\cos^2{(2\psi)}}{16(g_a^++1)^2(g_b^++1)^2\Delta^2}+\frac{g_a^-g_b^-(1-2s'^2)^2\sin^2{(2\psi)}}{16(g_a^++1)^2(g_b^++1)^2}\\&+\frac{E(1-2s'^2)^2\sin{(4\psi)}}{32(g_a^++1)^2(g_b^++1)^2\Delta}(g_b^--g_a^-)
	\end{align}
	For $R_2(k)$, we obtain in a similar fashion
	\begin{align} 	\alpha_1(s',t',\psi) &= \frac{(s't'+\sqrt{1+s'^2t'^2-2s'^2})^2}{8(g_a^++1)(g_b^+1)},\\
		\alpha_2(s',t',\psi)&\nonumber\nonumber=\frac{(2s'^2-1)(s't'+\sqrt{1+s'^2t'^2-2})}{16\sqrt{2}(g_a^++1)^2(g_b^++1)^2\Delta}\\&\nonumber\times 
		(iE(\exp{(2\psi i)}-\exp{(-2\psi i)})(2+g_a^++g_b^+)+(\exp{(2\psi i)}\\&+\exp{(-2\psi i)})(g_a^-(g_b^++1)-g_b^-(g_a^++1))\Delta),\\
		\alpha_3(s',t',\psi) &\nonumber= \frac{g_a^+g_b^+-g_a^+-g_b^+-3}{16(g_a^++1)^2(g_b^++1)^2}\Biggl(1+2s'^2(t'^2-2)+2s'^4(2-2t'^2+t'^4)\\\nonumber&+s't'\sqrt{1+s'^2(t'^2-2)}+2s'^3t'(t'^2-1)\sqrt{1+s'^2(t'^2-2)}\Biggr)\\\nonumber&+\frac{E^2(1-2s'^2)^2\sin^2{(2\psi)}}{16(g_a^++1)^2(g_b^++1)^2\Delta^2}-\frac{g_a^-g_b^-(1-2s'^2)^2\cos^2{(2\psi)}}{16(g_a^++1)^2(g_b^++1)^2}\\&+\frac{E(1-2s'^2)^2\sin{(4\psi)}}{32(g_a^++1)^2(g_b^++1)^2\Delta}(g_b^--g_a^-)
	\end{align}
	At this point we already observe that $\alpha_1(s',t',\psi)$ does not depend on $g_c^-,\ E,\ \Delta$ or $\psi$, which is also the case for the accompanying prefactors of the leading--order term. The asymptote deeply in the Ericson regime therefore solely depends on $g_a^+,\ g_b^+$ and $\Xi$.
	Setting the functions $\alpha_j$ in the approximations $J_\nu^{(l)}(k\omega),\ J_\nu^{(sl)}(k\omega), \ J_\nu^{(ssl)}(k\omega)$ and approximating their respective prefactors by a Taylor expansion in $p'$, we get the asymptotic form of the characteristic functions $R_1(k)$, $R_2(k)$. By doing so, we find that the prefactors also carry the exponential terms $\exp{(2\psi i l)}, \ l\in \mathbb{Z}.$
	To find the Gaussian asymptote, it is sufficient to only keep the leading order expansion of the Bessel functions multiplied with the leading--order term of the corresponding prefactor.  Calculating the phase integral then merely consists of picking out the contribution where the exponential has argument $0$. Then we carry out the $p'$--integral as a Laplace transform. This amounts to calculating integrals of the form 
	\begin{align}
		\int\limits_{0}^{\infty}dp'\exp{\left(-\frac{\pi\Xi p'}{2}\right)} p'^j J_{\nu}(2k\sqrt{\alpha_1(s',t',\psi)p'})= \mathcal{L}\left[p'^j J_{\nu}(2k\sqrt{\alpha_1(s',t',\psi)p'})\right]\left(\frac{\pi\Xi}{2}\right),
	\end{align}
	which is done by using the series representations of the Bessel functions and interchanging integration and summation. 
	The steps mentioned are, in principle, elementary, but extremely lengthy. We therefore refrain from presenting them here explicitly and continue by solving the remaining two--dimensional integral of the leading--order approximation over the disconnected part.
	\subsection{Derivation of the universal Gaussian}
 Analogously to the symplectic case, we make a substitution of the form
	\begin{align}\label{eq:coordinates}
		s' &= \sqrt{\frac{1-\sqrt{y(1+x(1-y))}}{2}},\ t' = \frac{\sqrt{1+x(1-y)}-\sqrt{y}}{\sqrt{2(1-\sqrt{y(1+x(1-y))})}},\\
		\left|\frac{\partial(s',t')}{\partial{(x,y)}}\right|&=\frac{(y-1)(\sqrt{y}+\sqrt{1+x(1-y))}}{16((1-x(y-1))y-\sqrt{y(1+x(1-y))})}.
	\end{align}
	This maps the integration region onto the rectangle $x\in [-1,1],\ y\in [0,1]$, as in the symplectic case. This also reintroduces the coordinate singularity at $x=y=1$ in such a way that we can explicitly see that it is integrable.
	The leading--order asymptote of the characteristic function is then written as 
	\begin{align}\label{eq:Rsk}
		R_s^{(l)}(k)& \nonumber=1+\int\limits_{-1}^{1}dx \int\limits_{0}^{1}dy\ k^2(1-y)(-1+x+y-xy)\\\nonumber&\Biggl[\exp{\left(-\frac{k^2y}{2(g_a^++1)(g_b^++1)(2\pi\Xi)}\right)} (k^4(y-1)y(xy-1)\\\nonumber&+2(g_a^++1)(g_b^++1)k^2(y-1)(2\pi\Xi)+4(g_a^++1)^2(g_b^++1)^2(2\pi\Xi)^2)\\\nonumber&+\exp{\left(-\frac{k^2(1-x(y-1))}{2(g_a^++1)(g_b^++1)(2\pi\Xi)}\right)}(k^4(1+x-xy+(1-x(y-1))^2y)+2(g_a^++1)(g_b^++1)k^2x(1-y)(2\pi\Xi)\\&+4(g_a^++1)^2(g_b^++1)^2(2\pi\Xi)^2\Biggr](16(g_a^++1)^3(g_b^++1)^3(1-y)^2(1-xy)^2(2\pi\Xi)^3)^{-1}.
	\end{align}
	The $y$-integral is done in terms of simple functions
	and yields 
	\begin{align}
		&\int\limits_{0}^{1}dy\ k^2(1-y)(-1+x+y-xy)\Biggl[\exp{\left(-\frac{k^2y}{2(g_a^++1)(g_b^++1)(2\pi\Xi)}\right)} \\\nonumber&\times(k^4(y-1)y(xy-1)+2(g_a^++1)(g_b^++1)k^2(y-1)(2\pi\Xi)\\\nonumber&+4 (g_a^++1)^2(g_b^++1)^2(2\pi\Xi)^2)\\\nonumber&+\exp{\left(-\frac{k^2(1-x(y-1))}{2(g_a^++1)(g_b^++1)(2\pi\Xi)}\right)}(k^4(1+x-xy+(1-x(y-1))^2y)-2(g_a^++1)(g_b^++1)k^2x(1-y)(2\pi\Xi)\\&+4(g_a^++1)^2(g_b^++1)^2(2\pi\Xi)^2\Biggr](16(g_a^++1)^3(g_b^++1)^3(1-y)^2(1-xy)^2(2\pi\Xi)^3)^{-1}\\\nonumber&
		\\\nonumber&=-k^2 \Biggl[2\exp{\left(-\frac{k^2}{2(g_a^++1)(g_b^++1)(2\pi\Xi)}\right)}(x^2+1)(g_a^++1)(g_b^++1)(2\pi\Xi)\\\nonumber&+\exp{\left(-\frac{k^2(x+1)}{2(g_a^++1)(g_b^++1)(2\pi\Xi)}\right)} (x^2-1)(k^2x+2(g_a^++1)(g_b^++1)(2\pi\Xi)\Biggr)\\& \times(8(g_a^++1)^2(g_b^++1)^2x^2(2\pi\Xi)^2)^{-1}.
	\end{align}
	The remaining integration is done similarly and results in 
	\begin{align}
		-k^2 \nonumber&\int\limits_{-1}^{1}dx\Biggl[2\exp{\left(-\frac{k^2}{2(g_a^++1)(g_b^++1)(2\pi\Xi)}\right)}(x^2+1)(g_a^++1)(g_b^++1)(2\pi\Xi)\\\nonumber&+\exp{\left(-\frac{k^2(x+1)}{2(g_a^++1)(g_b^++1)(2\pi\Xi)}\right)} (x^2-1)(k^2x+2(g_a^++1)(g_b^++1)(2\pi\Xi))\Biggr]\\\nonumber& \times(8(g_a^++1)^2(g_b^++1)^2x^2(2\pi\Xi)^2)^{-1}
	\\&= \exp{\left(-\frac{k^2}{(g_a^++1)(g_b^++1)(2\pi\Xi)}\right)}-1.
	\end{align}
	The contribution $-1$ from the lower integral boundary cancels the Efetov-Wegner-contribution, which is given by $1$ and ensures the normalization of the probability distribution. Using the fundamental theorem of calculus is justified, since in the small--$x$ limit the Taylor expansion of the exponential function gives
	\begin{align}
		\Biggl(1&-\exp{\Biggl(-\frac{k^2x}{2(g_a^++1)(g_b^++1)(2\pi\Xi)}\Biggr)}\Biggr) x^{-1} \\\nonumber&\simeq x^{-1}-x^{-1}+\frac{k^2x}{2(g_a^++1)(g_b^++1)(2\pi\Xi) x } +\mathcal{O}(x) = \frac{k^2}{2(g_a^++1)(g_b^++1)(2\pi\Xi) }+\mathcal{O}(x). 
	\end{align}
	Therefore, we have 
	\begin{align}
		R_s^{(l)}(k)= \exp{\left(-\frac{k^2}{(g_a^++1)(g_b^++1)(2\pi\Xi)}\right)}
	\end{align}
	as leading--order asymptote of the characteristic function. Performing the inverse Fourier transform then yields Gaussian behavior, 
	\begin{align}
		P_s^{(l)}(x_s)=\mathcal{F}^{-1}[R_s^{(l)}(k)](x_s) =\sqrt{\frac{ \Xi(g_a^++1)(g_b^++1)}{2}}\exp{\Biggl(-\frac{\pi \Xi(g_a^++1)(g_b^++1)x_s^2}{2}\Biggr)} \ .
	\end{align}
	Rescaling according to $\xi_s = \sqrt{\Xi}x_s$ yields the universal Gaussian
	\begin{align}
		P_s^{(l)}(\xi_s) = \sqrt{\frac{ (g_a^++1)(g_b^++1)}{2}}\exp{\Biggl(-\frac{\pi (g_a^++1)(g_b^++1)\xi_s^2}{2}\Biggr)}. 
	\end{align}
	This exact behavior was also found in the cases $\beta=2,\ 4.$ In the limit of large $\Xi$, the distributions of rescaled real-- and imaginary parts of $S$--matrix elements are the same in all three Wigner--Dyson classes, to which we refer as universality within a universality. As shown for $\beta=2,\ 4$, deviations from this Gaussian behavior arise already in the subleading--order asymptote to the characteristic functions. These terms then, generally, differ in the orthogonal, unitary and symplectic cases. To complete our analysis, we now present the subleading--order terms for the case of $\beta=1$.

\section{Higher order corrections}
	To get the desired subleading--order corrections to the Gaussian limit, we seperate the contributions that depend on the terms including $g_c^-$ and $E,\ \Delta$ and those that solely depend on $g_c^+,\ \Xi$. We start with the corrections to the characteristic functions $R_s(k)$ in Sec. III.A, before we get to the moments  $\overline{x_s^{2n}}$ in Sec. III.B and then to the distributions $P_s(\xi_s)$ in Sec. III.C.
	\subsection{Subleading--order corrections to $R_s(k)$}
	 We write 
	 \begin{align}
	 	R_s^{(sl)}(k) = 		R_s^{(sl, 1)}(k)+ R_s^{(sl, 2)}(k),
	 \end{align}
	 where $R_s^{(sl, 1)}(k)$ denotes the subleading--order correction which is independent from $g_c^-$ and $E,\ \Delta$ and $R_s^{(sl, 2)}(k)$ for the remaining correction. 
	 We then proceed by making use of the expansions of the Bessel functions and expanding all prefactors to subleading order in $p'$. We then collect the terms where $\exp{(2\psi i l)}=1$ and carry out the Laplace transforms. The remaining two--dimensional integrals are solved as in the leading--order approximation by making use of the coordinates given in \cref{eq:coordinates}.
This procedure, despite being extremely cumbersome, results in the correction term
	\begin{align}
		R_s^{(sl, 1)}(k)=\exp{\left(-\frac{k^2}{(g_a^++1)(g_b^++1)(2\pi\Xi)}\right)}\left(k^2\frac{2(2+g_a^++g_b^+)}{((g_a^++1)(g_b^++1)(2\pi\Xi))^2}+k^4\frac{(g_a^+g_b^+-g_a^+-g_b^+-3)}{((g_a^++1)(g_b^++1)(2\pi\Xi))^3}\right).
	\end{align}

	This correction is present for all values of $T_c$, $E$ and $g_c^-$. However, when all channels are perfectly coupled, \textit{e.g.} $T_c=1$, the above equation provides the only correction. From a mathematical viewpoint, this is the case when the phase $\phi$ of an $S$--matrix element $S_{ab} = r \exp{(i\phi)}$ is uniformly distributed, \textit{e.g.} $P_\phi(\phi)= 1/(2\pi)$. From the standpoint of the Ericson transition, this situation arises when one transits from the case of $M\gg1$ with $T_c=1$ to the case of only a few perfectly coupled channels. Under these circumstances, the above correction to the characteristic function will provide the relevant correction term. If $T_c\neq1$, however, one has to consider additional contributions depending on the phase of the $S$--matrix element, which is, in the orthogonal case, not uniformly distributed \cite{Matrixelemente}.
	As mentioned above, further asymptotic corrections depending on $E,\ \Delta$ and $g_c^-$ do not preserve the radial symmetry of the bivariate characteristic function or the joint distribution $P(\xi_1,\xi_2)$, equivalently. Therefore, the correction terms from the univariate characteristic functions,  $R_1^{(sl, 2)}(k)$ and $R_2^{(sl, 2)}(k)$, differ. However, since the underlying geometry of the manifold does not change, we can employ the same techniques to do the remaining integrals. By applying the same procedure mentioned above to the terms including $g_c^-,\ E,\ \Delta$, we arrive at
	\begin{align}
		R_{1}^{(sl, 2)}(k) &\nonumber= \exp{\left(-\frac{k^2}{(2\pi\Xi)(g_a^++1)(g_b^++1)}\right)}\frac{k^4}{(2\pi\Xi)^3(g_a^++1)^4(g_b^++1)^4}\\&\times\Biggl(\frac{E^2}{\Delta^2}(g_a^+-g_b^+)^2 + (g_a^-(g_b^++1)+g_b^-(g_a^++1))^2\Biggr), \\
		R_{2}^{(sl, 2)}(k)&\nonumber=\exp{\left(-\frac{k^2}{(2\pi\Xi)(g_a^++1)(g_b^++1)}\right)}\frac{k^4}{(2\pi\Xi)^3(g_a^++1)^4(g_b^++1)^4}\\&\times\Biggl(\frac{E^2}{\Delta^2}(2+g_a^++g_b^+)^2 + (g_a^-(g_b^++1)-g_b^-(g_a^++1))^2\Biggr).
	\end{align}
	As seen from the above equations, the polynomials accompanying the exponential are now different for $s=1,\ 2$. The characteristic functions are, up to subleading--order, given by 
	\begin{align}
	R_{s}(k) &\nonumber\simeq \exp{\left(-\frac{k^2}{2\pi\Xi(g_a^++1)(g_b^++1)}\right)}\\\nonumber&\times\Biggl(1+ k^2\frac{2(2+g_a^++g_b^+)}{(g_a^++1)^2(g_b^++1)^2(2\pi\Xi_\beta)^2}+k^4\Biggl(\frac{g_a^+g_b^+-g_b^+-g_b^+-3}{(g_a^++1)^3(g_b^++1)^3(2\pi\Xi_\beta)^3}\\&\nonumber+\Biggl(\delta_{1s}\frac{1}{((2\pi\Xi)^3(g_a^++1)^4(g_b^++1)^4)}\Biggl(\frac{E^2}{\Delta^2}(g_a^+-g_b^+)^2 + (g_a^-(g_b^++1)+g_b^-(g_a^++1))^2\Biggr)\\&+\delta_{2s}\frac{1}{((2\pi\Xi)^3(g_a^++1)^4(g_b^++1)^4)}\Biggl(\frac{E^2}{\Delta^2}(2+g_a^++g_b^+)^2 + (g_a^-(g_b^++1)-g_b^-(g_a^++1))^2\Biggr)\Biggr)\Biggr)\Biggr).
	\end{align}

The moments are immediately found by writing the exponential as a series and identifying the coefficient of $k^{2n}$. 
\subsection{Subleading--order corrections to $\overline{x_s^{2n}}$}
For the moments, we obtain
	\begin{align}
		\overline{x_1^{2n}} &\nonumber= \frac{(2n)!}{n!}\Biggl(\frac{1}{((2\pi\Xi)(g_a^++1)(g_b^++1))^n}\\&\nonumber-  \frac{2(2+g_a^++g_b^+)}{((2\pi\Xi)(g_a^++1)(g_b^++1))^2}\frac{n}{((2\pi\Xi)(g_a^++1)(g_b^++1))^{n-1}}\\&\nonumber+\Biggl(\frac{g_a^+g_b^+-g_a^+-g_b^+-3}{((2\pi\Xi)(g_a^++1)(g_b^++1))^3}+\frac{1}{((2\pi\Xi)^3(g_a^++1)^4(g_b^++1)^4)}\\&\nonumber\times\Biggl(\frac{E^2}{\Delta^2}(g_a^+-g_b^+)^2 + (g_a^-(g_b^++1)+g_b^-(g_a^++1))^2\Biggr)\Biggr)n(n-1)\\&\times \frac{1}{((2\pi\Xi)(g_a^++1)(g_b^++1))^{n-2}}\Biggr)
	\end{align}
	and 
	\begin{align}
		\overline{x_2^{2n}} &\nonumber= \frac{(2n)!}{n!}\Biggl(\frac{1}{((2\pi\Xi)(g_a^++1)(g_b^++1))^n}\\&\nonumber-  \frac{2(2+g_a^++g_b^+)}{((2\pi\Xi)(g_a^++1)(g_b^++1))^2}\frac{n}{((2\pi\Xi)(g_a^++1)(g_b^++1))^{n-1}}\\&\nonumber+\Biggl(\frac{g_a^+g_b^+-g_a^+-g_b^+-3}{((2\pi\Xi)(g_a^++1)(g_b^++1))^3}+\frac{1}{((2\pi\Xi)^3(g_a^++1)^4(g_b^++1)^4)}\\&\nonumber\times\Biggl(\frac{E^2}{\Delta^2}(2+g_a^++g_b^+)^2 + (g_a^-(g_b^++1)-g_b^-(g_a^++1))^2\Biggr)\Biggr)n(n-1)\\&\times \frac{1}{((2\pi\Xi)(g_a^++1)(g_b^++1))^{n-2}}\Biggr).
	\end{align}
	 The moments therefore separate into a contribution that is only dependent on $T_c$ and $\Xi$ that is equal for $s=1,\ 2$ and a contribution that also depends on $g_c^-,E,\Delta$. The latter encodes the deviations in the distributions of $x_s$.
	To find the associated distributions $P_s(\xi_s)$, $\xi_s = \sqrt{\Xi}x_s $, we calculate the inverse Fourier transform of the asymptotic expressions.
	\subsection{Subleading--order corrections to $P_s(\xi_s)$} For the distributions $P_s(\xi_s)$, we find
	\begin{align}
		P_s(\xi_s) &= \frac{1}{\sqrt{\Xi}}\mathcal{F}^{-1}[R_1^{(l)}(k) + R_1^{(sl, 1)}(k) +R_1^{(sl, 2)}(k) ]\left(\frac{\xi_s}{\sqrt{\Xi}}\right)\\&=P_s^{(l)}(\xi_s)+P_s^{(sl,1)}(\xi_s)+P_s^{(sl,2)}(\xi_s)
		\end{align}
		where 
		\begin{align}
		P_s^{(l)}(\xi_s)&=\sqrt{\frac{ (g_a^++1)(g_b^++1)}{2}}\exp{\Biggl(-\frac{\pi (g_a^++1)(g_b^++1)\xi_s^2}{2}\Biggr)}, \\	P_s^{(sl,1)}(\xi_s)&\nonumber=\sqrt{\frac{ (g_a^++1)(g_b^++1)}{2}}\exp{\Biggl(-\frac{\pi (g_a^++1)(g_b^++1)\xi_s^2}{2}\Biggr)}\Biggl(\frac{3(g_a^+g_b^+-g_a^+-g_b^+-3)+4(2+g_a^++g_b^+)}{8(g_a^++1)(g_b^++1)\pi\Xi}\\&\nonumber
		-\frac{6(g_a^++1)(g_b^++1)(g_a^+g_b^+-g_a^+-g_b^+-3)+4(g_a^++1)(g_b^++1)(2+g_a^++g_b^+)}{8\Xi(g_a^++1)(g_b^++1)}\xi_s^2\\&+\frac{(g_a^++1)(g_b^++1)(g_a^+g_b^+-g_a^+-g_b^+-3)\pi}{8\Xi}\xi_s^4\Biggr),\\
		P_s^{(sl,2)}(\xi_s)&\nonumber=\delta_{s1}\sqrt{\frac{ (g_a^++1)(g_b^++1)}{2}}\exp{\Biggl(-\frac{\pi (g_a^++1)(g_b^++1)\xi_s^2}{2}\Biggr)}\\&\nonumber\times\Biggl(
		\frac{3(E^2(g_a^+-g_b^+)^2+(g_a^-(g_b^++1)+g_b^-(g_a^++1))^2\Delta^2)}{8(g_a^++1)^2(g_b^++1)^2\pi\Delta^2\Xi}\\&\nonumber-	\frac{3(E^2(g_a^+-g_b^+)^2+(g_a^-(g_b^++1)+g_b^-(g_a^++1))^2\Delta^2)}{4(g_a^++1)(g_b^++1)\Delta^2\Xi}\xi_s^2\\&\nonumber+\frac{(E^2(g_a^+-g_b^+)^2+(g_a^-(g_b^++1)+g_b^-(g_a^++1))^2\Delta^2)}{8(g_a^++1)(g_b^++1)\Delta^2\Xi}\pi\xi_s^4\Biggr)\\&\nonumber+\delta_{s2}\sqrt{\frac{ (g_a^++1)(g_b^++1)}{2}}\exp{\Biggl(-\frac{\pi (g_a^++1)(g_b^++1)\xi_s^2}{2}\Biggr)}\\&\nonumber\times\Biggl(
		\frac{3(E^2(2+g_a^++g_b^+)^2+(g_a^-(g_b^++1)-g_b^-(g_a^++1))^2\Delta^2)}{8(g_a^++1)^2(g_b^++1)^2\pi\Delta^2\Xi}\\&\nonumber-	\frac{3(E^2(2+g_a^++g_b^+)^2+(g_a^-(g_b^++1)-g_b^-(g_a^++1))^2\Delta^2)}{4(g_a^++1)(g_b^++1)\Delta^2\Xi}\xi_s^2\\&+\frac{(E^2(2+g_a^++g_b^+)^2+(g_a^-(g_b^++1)-g_b^-(g_a^++1))^2\Delta^2)}{8(g_a^++1)(g_b^++1)\Delta^2\Xi}\pi\xi_s^4
		\Biggr).	
	\end{align}
	Thus, at $\xi_s=0$, we find a shift to the Gaussian maximum in the Ericson regime. Especially, we have $P_1(0)\neq P_2(0)$. At $\xi_s=0$, we arrive at
	\begin{align}
		P_1(0) &\nonumber= \sqrt{\frac{ (g_a^++1)(g_b^++1)}{2}}\exp{\Biggl(-\frac{\pi (g_a^++1)(g_b^++1)\xi_1^2}{2}\Biggr)}\\&\nonumber\times\Biggl(1+\frac{3(g_a^+g_b^+-g_a^+-g_b^+-3)+4(2+g_a^++g_b^+)}{8(g_a^++1)(g_b^++1)\pi\Xi}+\frac{3(E^2(g_a^+-g_b^+)^2+(g_a^-(g_b^++1)+g_b^-(g_a^++1))^2\Delta^2)}{8(g_a^++1)^2(g_b^++1)^2\pi\Delta^2\Xi}\Biggr)\\&+\mathcal{O}(\Xi^{-2}),\\
		P_2(0)&\nonumber=\sqrt{\frac{ (g_a^++1)(g_b^++1)}{2}}\exp{\Biggl(-\frac{\pi (g_a^++1)(g_b^++1)\xi_1^2}{2}\Biggr)}\\&\nonumber\times\Biggl(1+\frac{3(g_a^+g_b^+-g_a^+-g_b^+-3)+4(2+g_a^++g_b^+)}{8(g_a^++1)(g_b^++1)\pi\Xi}+\frac{3(E^2(2+g_a^++g_b^+)^2+(g_a^-(g_b^++1)-g_b^-(g_a^++1))^2\Delta^2)}{8(g_a^++1)^2(g_b^++1)^2\pi\Delta^2\Xi}\Biggr)\\&+\mathcal{O}(\Xi^{-2})
	\end{align}
	The maximum of the distributions therefore differs by the contributions dependent on $g_c^-,\ E,\ \text{and}\  \Delta$.

		\section{Cross--section distribution}
	In many experiments only the cross sections are available via measurement. Therefore, having the distribution of cross--sections at hand is extremely important from an experimental point of view. Thus, we extend our analysis to the distribution of cross--sections, which was derived in \cite{Wirkungsquerschnitte}.  There, the bivariate characteristic function $R(\boldsymbol{k}) = R(k_1,k_2), \ \boldsymbol{k}=k_1+i k_2$ was given as an integral over singular values by
	
	\begin{align}\label{eq:bichar}
R(k_1,k_2)=1+\frac{1}{8\pi}\int\limits_{-1}^{1}d\lambda_0\int\limits_{1}^{\infty}d\lambda_1 \int\limits_{1}^{\infty} d\lambda_2\int\limits_{0}^{2\pi}d\psi J(\lambda_0,\lambda_1,\lambda_2)\mathcal{F}_O(\lambda_0,\lambda_1,\lambda_2)(\kappa_1+\kappa_2+\kappa_3+\kappa_4),
\end{align}
	where 
	\begin{align}\label{eq:Berezinian}
J(\lambda_0,\lambda_1,\lambda_2) = \frac{(1-\lambda_0^2)|\lambda_1-\lambda_2|}{2(\lambda_1^2-1)^{1/2}(\lambda_2^2-1)^{1/2}(\lambda_1-\lambda_0)^2(\lambda_2-\lambda_0)^2}
	\end{align}
	and
\begin{align}\label{eq:cf}
	\mathcal{F}_0(\lambda_0,\lambda_1,\lambda_2)&=\prod\limits_{c=1}^{M}\frac{g_c^++\lambda_0}{(g_c^++\lambda_1)^{1/2}(g_c^++\lambda_2)^{1/2}}.
\end{align}
The $\kappa_j$ in \cref{eq:bichar} now contain Bessel functions and algebraic functions. The full integrand of \cref{eq:bichar} is given in Appendix E.
As the Berezinian given in \cref{eq:Berezinian} is the same as for the univariate characteristic functions, the methods presented here to lift the singularity also carry over to the cross--section distribution. The channel--factor in \cref{eq:cf} also remains  the same, such that it is approximated by the exponential in \cref
{eq:exponential}. The generation of the asymptotic expansion therefore proceeds completely analogous to the case of the univariate characteristic functions, with minor modifications to the multinomial expansion of the Bessel functions.
	
After applying the procedure as outlined, we find that the bivariate characteristic function is, up to subleading--order, given by
	\begin{align}	R(k_1,k_2)&\simeq R^{(l)}(k_1,k_2)+R^{(sl,1)}(k_1,k_2)+R^{(sl,2)}(k_1,k_2),\\
		R^{(l)}(k_1,k_2)&= \exp{\left(-\frac{k_1^2+k_2^2}{(g_a^++1)(g_b^++1)(2\pi\Xi)}\right)},\\
		R^{(sl,1)}(k_1,k_2)&\nonumber=\frac{k_1^2+k_2^2}{(g_a^++1)^4(g_b^++1)^4(2\pi\Xi)^3}\exp{\left(-\frac{k_1^2+k_2^2}{(g_a^++1)(g_b^++1)(2\pi\Xi)}\right)}\\\nonumber&\nonumber\times\Biggl( (g_a^+g_b^+-g_a^+-g_b^+-3)(g_a^++1)(g_b^++1)(k_1^2+k_2^2)\\&+2(g_a^++1)^2(g_b^++1)^2(2+g_a^++g_b^+)(2\pi\Xi)\Biggr),\\
			R^{(sl, 2)}(k_1,k_2)&\nonumber=	\exp{\left(-\frac{k_1^2+k_2^2}{(g_a^++1)(g_b^++1)2\pi\Xi}\right)}\\&\nonumber\times\Biggl( (g_a^-(g_b^++1) +g_b^-(g_a^++1))^2k_1^2+ (g_a^-(g_b^++1) -g_b^-(g_a^++1))^2k_2^2 \\&\nonumber+\frac{E^2}{\Delta^2}((g_a^+-g_b^+)^2k_1^2 +(2+g_a^++g_b^+)^2k_2^2) + \frac{4E}{\Delta}(g_a^++1)(g_b^++1)(g_a^-+g_b^-)k_1k_2 \Biggr)\\&\times \frac{k_1^2+k_2^2}{(g_a^++1)^4(g_b^++1)^4(2\pi\Xi)^3}
		\label{eq:bivariateGOE}.\end{align}
	As seen from \cref{eq:bivariateGOE}, the bivariate characteristic function is not radially symmetric. The symmetry is broken by the subleading--order terms coupled to $g_a^-,g_b^-, E$. If one considers the case $T_c=1$ for all channels, which corresponds to the COE, the radial symmetry is preserved, since we then have 
	\begin{align}
		R^{(sl,1)}(k_1,k_2)&=\frac{k_1^2+k_2^2}{(g_a^++1)^4(g_b^++1)^4(2\pi\Xi)^3}\exp{\left(-\frac{k_1^2+k_2^2}{(g_a^++1)(g_b^++1)(2\pi\Xi)}\right)}\\\nonumber&\times\Biggl( (g_a^+g_b^+-g_a^+-g_b^+-3)(g_a^++1)(g_b^++1)(k_1^2+k_2^2)\\\nonumber&+2(g_a^++1)^2(g_b^++1)^2(2+g_a^++g_b^+)(2\pi\Xi)\Biggr)
	\end{align}
	as the only correction.
	The term breaking the radial symmetry is
	\begin{align}
		R^{(sl, 2)}(k_1,k_2)&\nonumber=	\exp{\left(-\frac{k_1^2+k_2^2}{(g_a^++1)(g_b^++1)2\pi\Xi}\right)}\\&\nonumber\times\Biggl( (g_a^-(g_b^++1) +g_b^-(g_a^++1))^2k_1^2+ (g_a^-(g_b^++1) -g_b^-(g_a^++1))^2k_2^2 \\&\nonumber+\frac{E^2}{\Delta^2}((g_a^+-g_b^+)^2k_1^2 +(2+g_a^++g_b^+)^2k_2^2) + \frac{4E}{\Delta}(g_a^++1)(g_b^++1)(g_a^-+g_b^-)k_1k_2 \Biggr)\\&\times \frac{k_1^2+k_2^2}{(g_a^++1)^4(g_b^++1)^4(2\pi\Xi)^3}.
	\end{align}
	The joint distribution $P(x_1,x_2)$ is found by taking the inverse two--dimensional Fourier transform,
	\begin{align}
		P(x_1,x_2) = \frac{1}{4\pi^2}\int\limits_{-\infty}^{\infty}dk_1\int\limits_{-\infty}^{\infty}dk_2\ \exp{(i k_1 x_1 + i k_2 x_2)}R(k_1,k_2),
	\end{align}
	while the cross--section distribution is calculated via the Bessel transform 
	\begin{align}
		p(\sigma_{ab}) = \frac{1}{4\pi}\int\limits_{-\infty}^{\infty}\int\limits_{-\infty}^{\infty}dk_1dk_2 \ R(k_1,k_2)J_0\left(\sqrt{(k_1^2+k_2^2)\sigma_{ab}}\right).
	\end{align}
	By expressing $R(k_1,k_2)$ in polar coordinates, \textit{e.g.} $k_1 = K \cos \phi,\ k_2 = K \sin \phi$,  we find 
	\begin{align}
		p(\sigma_{ab})\nonumber&\simeq \frac{1}{4\pi}\int\limits_{0}^{\infty}\int\limits_{0}^{2\pi}dK d\phi \ K (R^{(l)}(K \cos \phi,K \sin \phi) + R^{(sl,1)}(K \cos \phi,K \sin \phi)+ R^{(sl,2)}(K \cos \phi,K \sin \phi)) J_0\left( K \sqrt{\sigma_{ab}}\right)\\&=\frac{1}{4\pi}\mathcal{H}_0\left[\int\limits_{0}^{2\pi}d\phi(R^{(l)}(K \cos \phi,K \sin \phi) + R^{(sl,1)}(K \cos \phi,K \sin \phi)+ R^{(sl,2)}(K \cos \phi,K \sin \phi))\right](\sqrt{\sigma_{ab}}).
	\end{align}
	Therefore, analogously to the unitary and symplectic cases, the cross--section distribution is expressed in terms of a Hankel transform. Due to the non--trivial angular dependency in the orthogonal case, the integration does not amount to a simple factor of $2\pi$ in contrast to the cases $\beta=2, 4$. 
	
	The remaining integral is the Hankel transform of a Gaussian-like function and a polynomial of even degree, which is easily calculated. After doing so, we rescale the cross--section with respect to its mean value in the Ericson regime, such that the leading--order term is an exponential decay. Finally, we arrive at the rescaled cross-section distribution 
	\begin{align}
		p(\widetilde{\sigma}_{ab}) &\simeq p^{(l)}(\widetilde{\sigma}_{ab})+p^{(sl,1)}(\widetilde{\sigma}_{ab})+p^{(sl,2)}(\widetilde{\sigma}_{ab}),\\
		p^{(l)}(\widetilde{\sigma}_{ab})&=\exp{(-\widetilde{\sigma}_{ab})},\\ 
			p^{(sl,1)}(\widetilde{\sigma}_{ab})&\nonumber=\frac{1}{2(g_a^++1)^2(g_b^++1)^2\pi\Xi}\exp{(-\widetilde{\sigma}_{ab})}\\&\times\Biggl(
			(g_a^++1)(g_b^++1)(-2+2 g_a^+g_b^++2(4+g_a^++g_b^+-2g_a^+g_b^+)\widetilde{\sigma}_{ab}+(g_a^+g_b^+-g_a^+-g_b^+-3)\widetilde{\sigma}_{ab}^2)\Biggr)
			\\
			p^{(sl,2)}(\widetilde{\sigma}_{ab})&\nonumber=\frac{1}{2(g_a^++1)^2(g_b^++1)^2\pi\Xi}\exp{(-\widetilde{\sigma}_{ab})}\Biggl(((g_a^++1)^2(g_b^-)^2 + (g_a^-)^2(g_b^++1)^2)(2-4\widetilde{\sigma}_{ab}+\widetilde{\sigma}_{ab}^2)\\&+ \frac{E^2}{\Delta^2}(2+g_a^+(g_a^++2)+g_b^+(2+g_b^+))(2-4\widetilde{\sigma}_{ab}+\widetilde{\sigma}_{ab}^2)
			\Biggr).
	\end{align}
	At $\widetilde{\sigma}_{ab}=0$, we find a shift, such that 
	\begin{align}
		p(0) &\nonumber= 1+ \frac{1}{2(g_a^++1)^2(g_b^++1)^2\pi\Xi}\Biggl(((g_a^++1)^2(g_b^-)^2 + (g_a^-)^2(g_b^++1)^2)2\\&+ \frac{E^2}{\Delta^2}(2+g_a^+(g_a^++2)+g_b^+(2+g_b^+))2+(g_a^++1)(g_b^++1)(-2+2 g_a^+g_b^+)\Biggr).
	\end{align}
	Especially, we observe that the correction term has no linear energy dependence, whereas the corrections dependent on $g_c^-$ and $E^2$ are still present.  
	\section{Comparison with numerical simulations}
\begin{figure}[H]
		\centering
		\captionsetup[subfigure]{labelformat=empty}
		\begin{subfigure}[t]{0.45\linewidth}
			\centering
			\includegraphics[width=\linewidth]{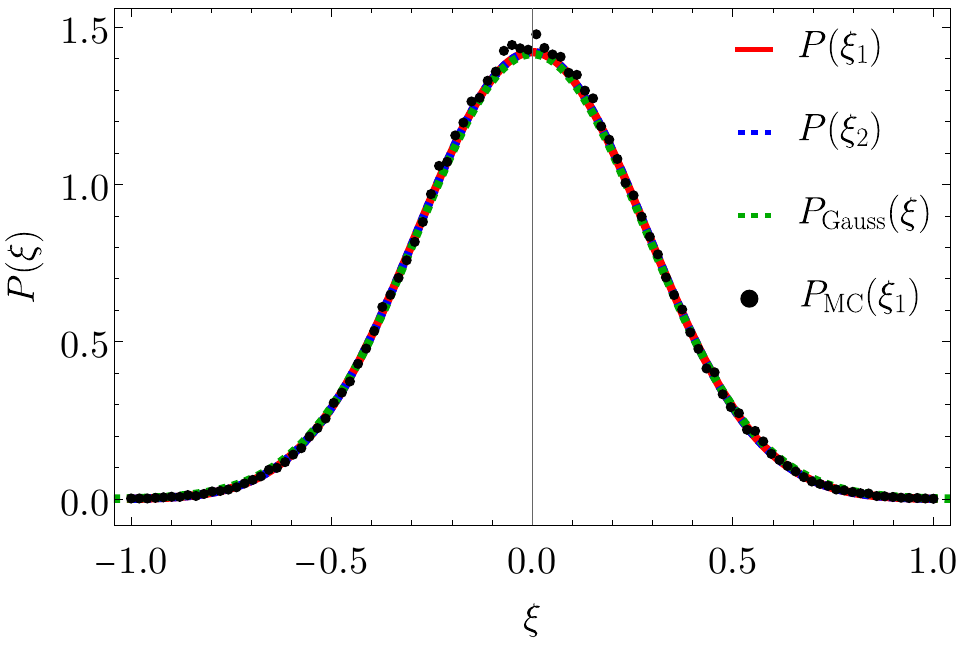} 
			\subcaption{}
		\end{subfigure}
		~
			\begin{subfigure}[t]{0.45\linewidth}
			\centering
			\includegraphics[width=\linewidth]{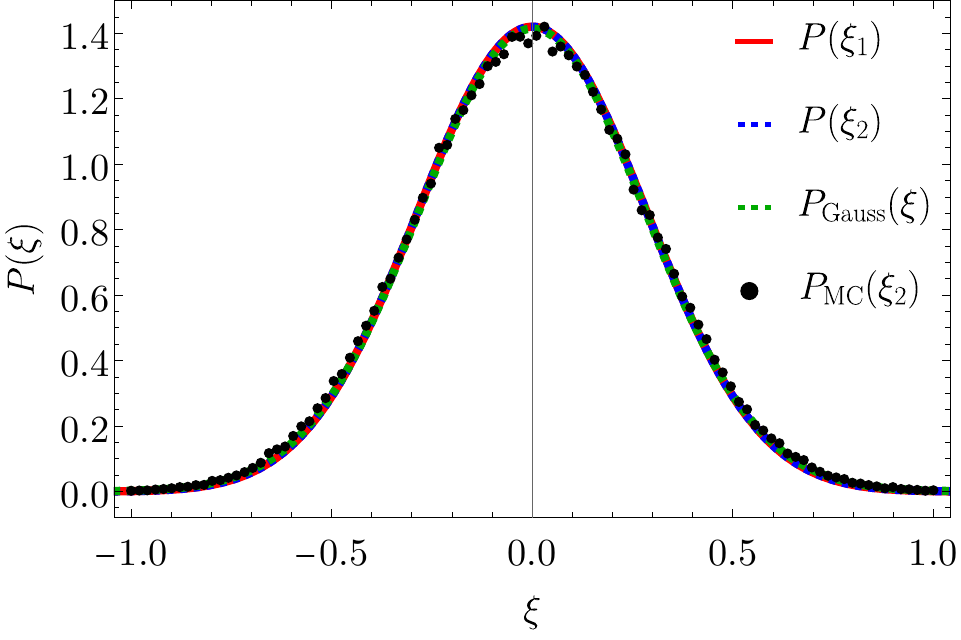} 
			\subcaption{}
		\end{subfigure}
		~
		\begin{subfigure}[t]{0.45\linewidth}
			\centering
			\includegraphics[width=1\linewidth]{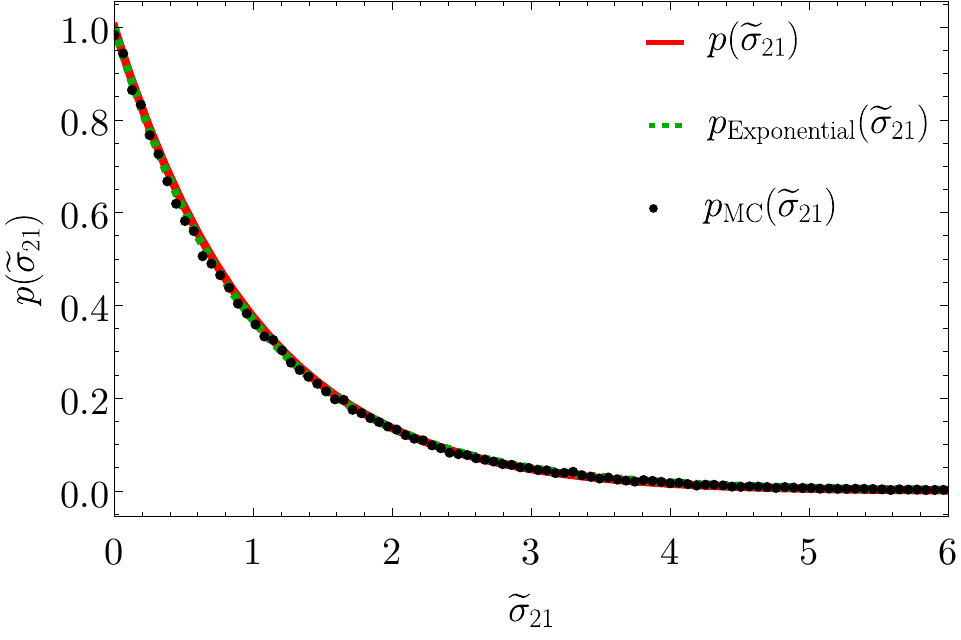} 
			\subcaption{}
		\end{subfigure}	
		~
		\begin{subfigure}[b]{0.45\linewidth}
			\centering
			\includegraphics[width=1\linewidth]{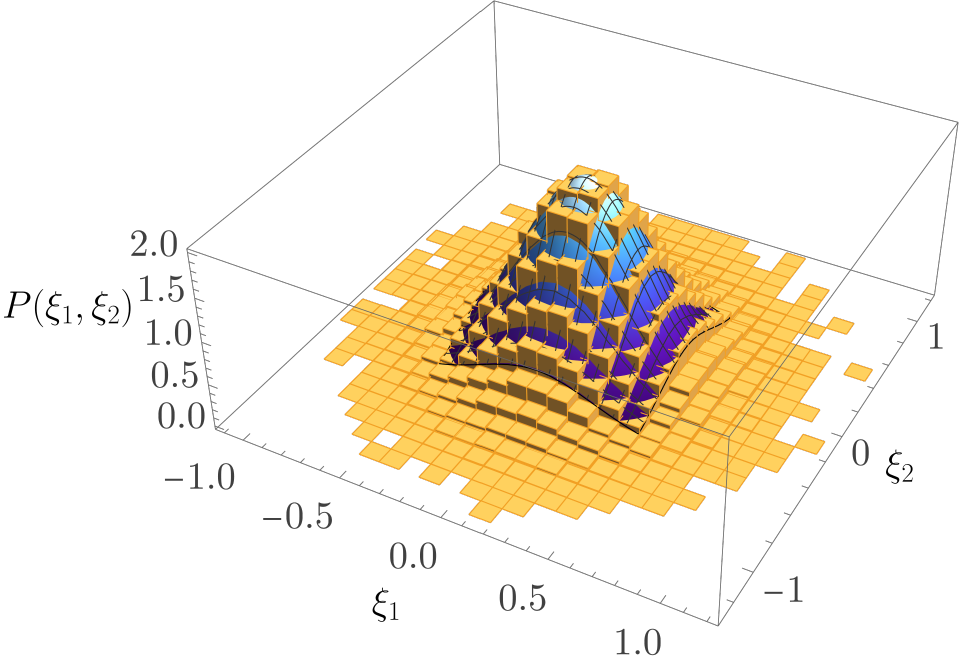} 
			\subcaption{}
		\end{subfigure}	
		~

		\caption{Numerical simulation and the distributions of rescaled scattering--matrix
			elements (top), cross--sections and joint distribution (bottom).
			$\Xi=9.23$ and $M=60$.}
		\label{fig:Corrections4}
	\end{figure}

	To further validate our analytical results, we compare with Monte Carlo simulations. We draw $100000$ matrices with $N=250$ from the GOE. To see the deviations of the distribution of real and imaginary part, we use $E=1, \nu=1$ and $T_1=T_2=0.579,\ \Xi=1.862$, $T_c=0.293,\ c\neq1,2$. This corresponds to $g_1^+=g_2^+=2.454,\ g_1^-=g_2^-=-2.165.$
	\begin{figure}[H]
		\centering
		\captionsetup[subfigure]{labelformat=empty}
		\begin{subfigure}[t]{0.45\linewidth}
			\centering
			\includegraphics[width=\linewidth]{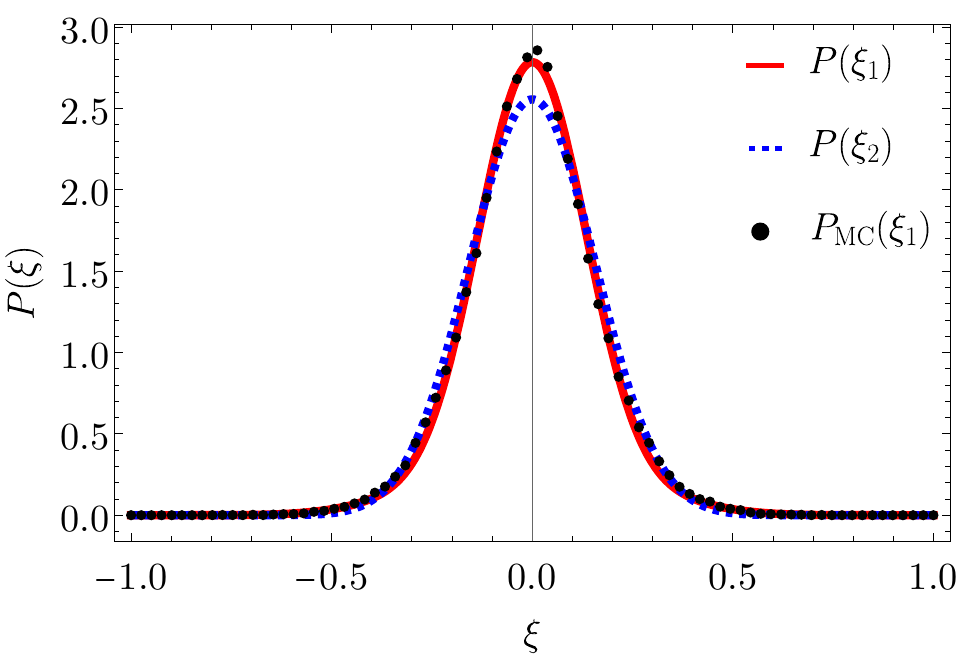} 
			\subcaption{}
		\end{subfigure}
		~
		\begin{subfigure}[t]{0.45\linewidth}
			\centering
			\includegraphics[width=\linewidth]{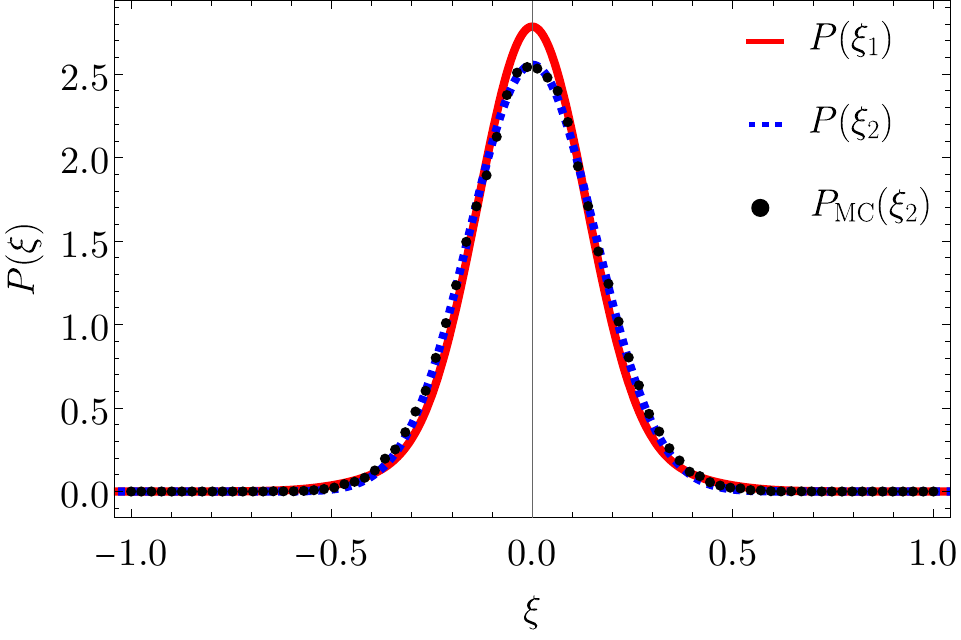} 
			\subcaption{}
		\end{subfigure}
		~
		\begin{subfigure}[t]{0.45\linewidth}
			\centering
			\includegraphics[width=\linewidth]{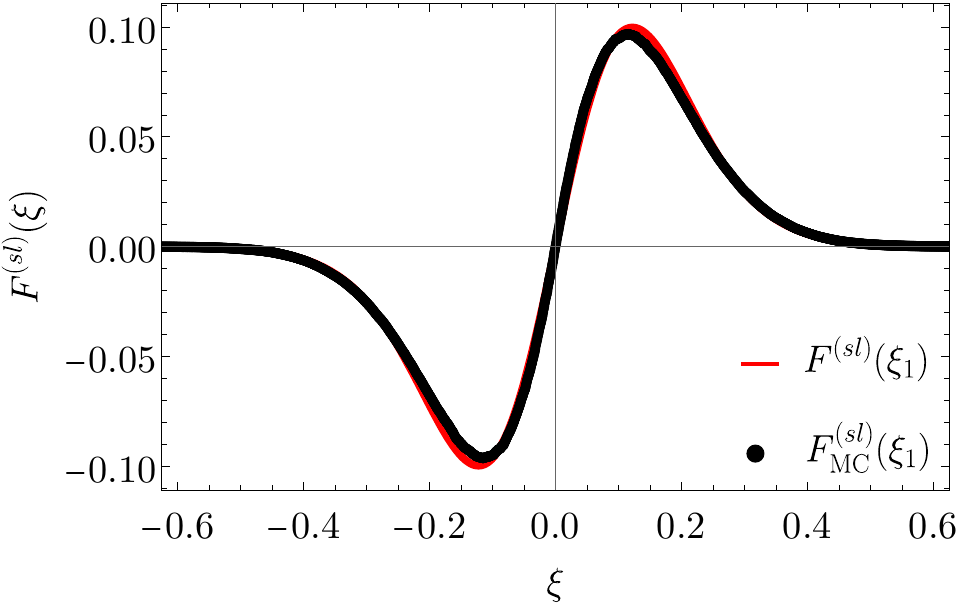} 
			\subcaption{}
		\end{subfigure}
		~
		\begin{subfigure}[t]{0.45\linewidth}
			\centering
			\includegraphics[width=1\linewidth]{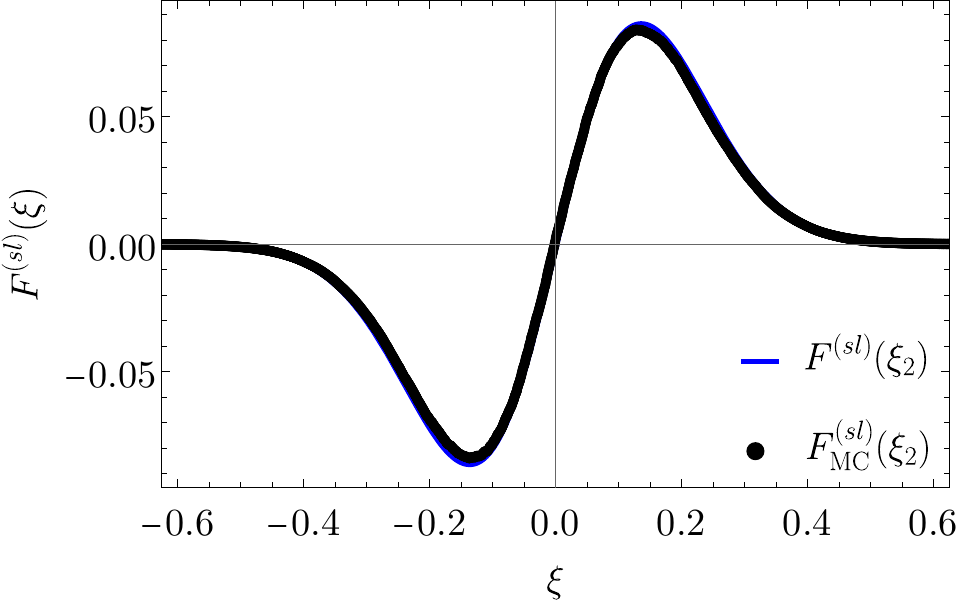} 
			\subcaption{}
		\end{subfigure}	
		~
		\begin{subfigure}[b]{0.45\linewidth}
			\centering
			\includegraphics[width=1\linewidth]{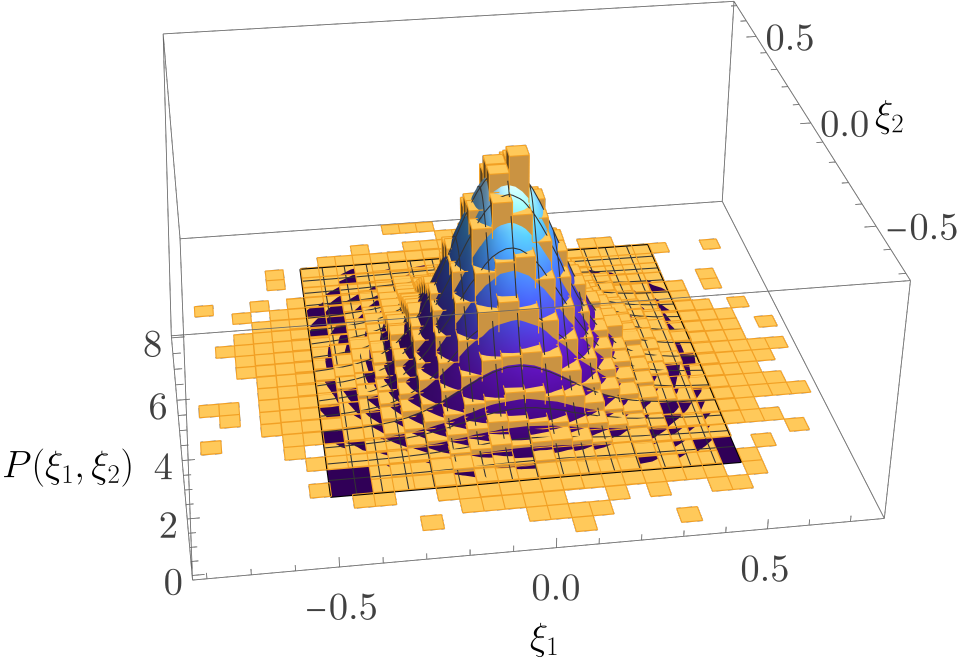} 
			\subcaption{}
		\end{subfigure}	
		\caption{Numerical simulation and the distributions of rescaled scattering matrix
			elements (top), full subleading correction for the CDF (mid) and joint distribution (bottom).
			$\Xi=1.862$ and $M=40$.}
		\label{fig:Corrections3}
	\end{figure}
	From \cref{fig:Corrections3} we see that $P_1(\xi_1)\neq P_2(\xi_2)$, which is caused by the correction term including $g_c^-, E$. The joint distribution $P(\xi_1, \xi_2)$ does not exhibit radial symmetry anymore. In \cref{fig:Corrections4} we find the predicted Gaussian behavior for $P_s(\xi_s)$ and exponential behavior for $p(\widetilde{\sigma}_{ab})$, where we use $M=60,\ \Xi=9.23, \ T_c=1$. 
	\section{Synopsis of the three Wigner-Dyson classes}
	Having the results for all three Wigner-Dyson classes at hand, we see that the Ericson regime gives rise to the same universal limit of the rescaled distributions in the case of $\Xi \gg1$, to which we refer as a universality within universality.
	This is astonishing, since the spin--symmetries and behavior unter time--reversal invariance are different in each class. Using the results presented above, we are able to write the asymptotes for the univariate characteristic functions, which we now denote by $R_{s\beta}(k)$ for comparison, in all three cases as
	\begin{align}
		R_{s\beta}(k) &\nonumber\simeq \exp{\left(-\frac{k^2}{(g_a^++1)(g_b^++1)(2\pi\Xi_\beta)}\right)}\\\nonumber&\times\Biggl(1+\chi_\beta k^2\frac{2(2+g_a^++g_b^+)}{(g_a^++1)^2(g_b^++1)^2(2\pi\Xi_\beta)^2}+k^4\Biggl(\frac{g_a^+g_b^+-g_b^+-g_b^+-3}{(g_a^++1)^3(g_b^++1)^3(2\pi\Xi_\beta)^3}+\delta_{1\beta}\\&\nonumber\times\Biggl(\delta_{1s}\frac{1}{((2\pi\Xi)^3(g_a^++1)^4(g_b^++1)^4)}\Biggl(\frac{E^2}{\Delta^2}(g_a^+-g_b^+)^2 + (g_a^-(g_b^++1)+g_b^-(g_a^++1))^2\Biggr)\\&+\delta_{2s}\frac{1}{((2\pi\Xi)^3(g_a^++1)^4(g_b^++1)^4)}\Biggl(\frac{E^2}{\Delta^2}(2+g_a^++g_b^+)^2 + (g_a^-(g_b^++1)-g_b^-(g_a^++1))^2\Biggr)\Biggr)\Biggr)\Biggr),\\
		\chi_\beta &= \begin{cases} \ \ 1,\  \beta = \ 1,\\
		\  \ 0, \ \beta=2,\\
			-1,\ \beta=4,
		\end{cases} \Xi_\beta = \begin{cases}\ \Xi, \ \beta = 1, 2\\
			2\Xi,\ \beta=4
		\end{cases}
	\end{align}
	The deviations from the Gaussian behavior are inferred from the univariate characteristic functions. Asymptotically, the characteristic functions are given by a Gaussian-like function times a polynomial of fourth order, where the squared term in the polynomial vanishes for $\beta=2$. Structurally, the cases $\beta=1$ and $\beta=4$ show similar asymptotic behavior in the quartic term when only the case of perfect coupling is considered. For non--perfect coupling, additional dependencies on the phase of the $S$--matrix arise in the orthogonal case. These or similar additional terms do not show up for $\beta=2,\ 4$, since there the phase is uniformly distributed. For $\beta=1,\ 4 $ the squared term within the polynomial, however, has a different sign, but the structural dependence on the transmission coefficients is the same. This is in accordance with the behavior of $S$--matrix correlation functions, where the unitary class lies between the orthogonal and symplectic case. If the radial symmetry is broken and we consider Hamiltonians from the GOE, an explicit energy dependency enters the asymptotes as well and subsequently yields deviations for the distributions of real and imaginary part.
	In view of the numerical simulations, we further emphasize that the transition to Ericson regime is occurring fast. The subleading--order correction is sufficient to capture non--Gaussian and non--exponential behavior of the respective distributions in all three ensembles. 
	\section{Conclusion}
	For systems with preserved time--reversal invariance and even spin, we analytically derived asymptotic expansions in the Ericson regime for the distribution of real and imaginary parts of scattering matrix elements, the cross--section distribution and moments. Lifting the coordinate singularity and making use of the separation into connected and disconnected parts of the saddle--point manifold, we exactly evaluated the integrals over the disconnected part. A cumbersome phase integral in the case of $\beta=1$ was solved by applying a multinomial expansion. The distribution of scattering--matrix elements results, upon a suitable rescaling within the respective symmetry class, in the same Gaussian asymptote in all three Wigner-Dyson classes, thereby revealing that the Ericson limit is also a universal limit arising from a universality. The subleading--order terms now show the transition from weakly to strongly overlapping resonances, where we explicitly see that the subleading--terms from outside the COE asymptotically break the radial symmetry of the bivariate characteristic function, and therefore of the joint distribution, too, in the case of $\beta=1$.
	\section*{Acknowledgments}
	We thank Nils Gluth and Barbara Dietz for fruitful discussions. The authors are grateful to the Deutsche Forschungsgemeinschaft (DFG, German Research Foundation) for support within the project Stochastic Quantum Scattering—New Tools, New Aspects, DFG Project No. 540160740. 
	\newpage 
\section*{Appendix}
	\subsection*{A --- Explicit definition of the coefficients $\kappa_{ij}$}
	In general, $\kappa_{ij}$ in the characteristic functions $R_s(k)$ are given by
	\begin{align}
		\kappa_{11} &= -\frac{9}{8} \{p_a^+ \sqrt{m}\}_+,\\\nonumber
		\kappa_{21}&= -\frac{1}{4}(128p_a^0p_b^0+14p_a^+ p_b^+ +32p_a^- p_b^-)\\&+\{3\exp(2\psi i)(p_a^-q_b^+-p_b^-r_a^+)\}_{-}+\{\exp{(-4\psi i)} q_a^-r_b^-)\}_+,\\
		\kappa_{22} &= -\frac{1}{4}\{(p_a^+ p_a^+ -4q_a^-r_a^-)m     \}_+,\\
		\kappa_{31}\nonumber&=\{2[(p_a^+ p_a^+ + q_a^-r_a^-)\sqrt{m} + 2(8p_a^0p_b^0+p_a^+p_b^+ +p_a^-p_b^-)\sqrt{l}](\exp(2\psi i)q_b^-+\exp(-2\psi i) r_b^-))\}_-\\\nonumber&+\{2[p_a^+p_b^-+4p_a^-p_b^+)\sqrt{m}+p_b^+p_b^-\sqrt{l}](\exp(-2\psi i)q_a^+ + \exp(2\psi i)r_a^+)\}_- \\\nonumber&+ \{[16p_a^0(2p_a^0p_b^+-3p_b^0p_a^+)-6p_a^+(q_a^+q_b^+ +r_a^+r_b^+)\\\nonumber&+2p_b^+(4q_a^+r_a^+-q_a^-r_a^-)-4p_a^-(p_a^+p_b^- - 2p_a^-p_b^+)+3p_a^+(q_a^-q_b^-+r_a^-r_b^--p_a^+p_b^+)\\\nonumber&+\left(\frac{1}{2}\exp(-4\psi i)\right)q_a^-(4p_a^+r_b^- - 3p_b^+q_a^-) + \left(\frac{1}{2}\exp(4\psi i)\right)r_a^-(4p_a^+q_b^--3p_b^+r_a^-)]\sqrt{m}\}_+\\\nonumber
		&+\Biggl\{\Biggl[\left(\frac{1}{2}\exp(-4\psi i)\right)q_a^-(2\exp(-2\psi i)q_a^-r_b^--8\exp(2\psi i)r_a^+r_b^+)\\&+\left(\frac{1}{2}\exp(4\psi i)\right)r_a^-(2\exp(2\psi i) q_b^-r_a^--8\exp(-2\psi i)q_a^+q_b^+)\Biggr]\sqrt{m}\Biggr\}_-,\\\nonumber
		\kappa_{32}&=\Biggl\{p_a^+\Biggl[(p_a^+p_a^++2q_a^-r_a^-)+\frac{3}{2}\Biggl(\exp(-4\psi i)q_a^-q_a^-+\exp(4\psi i)r_a^-r_a^-\Biggr)\Biggr]m^{3/2}\Biggr\}_+\\&+
		\{(2p_a^+p_a^++q_a^-r_a^-)(\exp(-2\psi i)q_a^-+\exp(2\psi i)r_a^-)m^{3/2}\}_-
	\end{align}
	and
	\begin{align}
		\kappa_{41}&= 32[2p_a^0p_a^0(p_b^--\exp(2\psi i) q_b^+)(p_b^- - \exp(-2\psi i)r_b^+)+2p_b^0p_b^0(p_a^-+\exp(-2\psi i)(p_a^-+\exp(2\psi i)r_a^+)\\\nonumber
		&+ p_a^0p_b^0((p_a^++\exp(-2\psi i)q_a^-)(p_b^+ - \exp(-2\psi i)r_b^-)+(p_a^++\exp(2\psi i)r_a^-)(p_b^+ - \exp(2\psi i) q_b^-))]\\\nonumber&+ 256p_a^0p_a^0p_b^0p_b^0+(p_a^+ + \exp(-2\psi i)q_a^-)^2(p_b^+ - \exp(-2\psi i)r_b^-)^2\\\nonumber&+(p_a^+ + \exp(2\psi i)r_a^-)^2(p_b^+-\exp(2\psi i)q_b^-)^2\\\nonumber&+ 4[(p_a^+ + \exp(-2\psi i) q_a^-)(p_b^+-\exp(2\psi i )q_b^-)-2(p_a^-+\exp(-2\psi i)q_a^+)(p_b^- - \exp(2\psi i)q_b^+)]\\
		&\times[(p_a^+ + \exp(2\psi i)r_a^-)(p_b^+ - \exp(-2\psi i)r_b^-)-2(p_a^- + \exp(2\psi i)r_a^+)(p_b^- - \exp(-2\psi i )r_b^+)].
	\end{align}
	Finally, we have
	\begin{align}
		\kappa_{42} &= -32p_a^0p_b^0[(p_a^+ + \exp(-2\psi i)q_a^-)(p_a^+ + \exp(2\psi i) r_a^-)m +(p_b^+ -\exp(2\psi i)q_b^-)(p_b^+ -\exp(-2\psi i)r_b^-)l]\\\nonumber& - 2[(p_a^++\exp(-2\psi i)q_a^-)(p_b^+ -\exp(2\psi i)q_b^-)-2(p_a^-+\exp(-2\psi i)q_a^+)(p_b^--\exp(2\psi i)q_b^+)]\\\nonumber&\times[(p_a^+ + \exp(2\psi i)r_a^-)^2m+(p_b^+ - \exp(-2\psi i )r_b^-)^2l]\\\nonumber&-2[(p_a^+ + \exp(2\psi i)r_a^-)(p_b^+ - \exp(-2\psi i)r_b^-) - 2(p_a^-+\exp(2\psi i )r_a^+)(p_b^--\exp(-2\psi i)r_b^+)]\\&\times[(p_a^++ \exp(-2\psi i )q_a^-)^2m+(p_b^+ - \exp(2\psi i)q_b^-)^2l],\\
		\kappa_{43}&= 
		(p_a^+ + \exp(-2\psi i )q_a^-)^2(p_a^+ + \exp(2\psi i )r_a^-)^2m^2+(p_b^+ -\exp(2\psi i)q_b^-)^2(p_b^+ -\exp(-2\psi i )r_b^-)^2l^2,
	\end{align}
	where
	\begin{align}
		\{\mathcal{E}(a,b,l,m,\psi)\}_\pm = \mathcal{E}(a,b,l,m,\psi)\pm\mathcal{E}(b,a,m,l,-\psi)
	\end{align}
	was used. In the leading--order approximation in the Ericson transition, the $\psi$-integral only yields a factor $2\pi$ and massively simplifies the coefficients to
	\begin{align}
		\kappa_{11} &= -\frac{9}{8} \{p_a^+ \sqrt{m}\}_+,\\\nonumber
		\kappa_{21}&= -\frac{1}{4}(128p_a^0p_b^0+14p_a^+ p_b^+ +32p_a^- p_b^-)\\\nonumber
		\kappa_{22} &= -\frac{1}{4}\{p_a^+ p_a^+ m     \}_+\\\nonumber
		\kappa_{31}&=\{[(16p_a^0(2p_a^0 p_b^+-3p_b^0p_a^+)\\\nonumber&-4p_a^-(p_a^+p_b^- - 2p_a^-p_b^+)+3p_a^+(-p_a^+p_b^+)]\sqrt{m}\}_+\\\nonumber
		\kappa_{32}&=\{p_a^+p_a^+p_a^+m^{3/2}\}_+\\\nonumber
		\kappa_{41}&= 32[2p_a^0p_a^0p_b^-p_b^- +2p_b^0p_b^0p_a^-p_a^-\\\nonumber
		&+ p_a^0p_b^0(p_a^+p_b^++p_a^+p_b^+)]
		+ 256p_a^0p_a^0p_b^0p_b^0+2(p_a^+)^2(p_b^+)^2\\\nonumber&+ 4[p_a^+p_b^+-2p_a^-p_b^- ]
		[p_a^+p_b^+ -2p_a^-p_b^-]\\\nonumber
		\kappa_{42} &= -32p_a^0p_b^0[(p_a^+)(p_a^+ )m +(p_b^+ )(p_b^+ )l] - 2[(p_a^+)(p_b^+ )-2(p_a^-)(p_b^-)]\\\nonumber&\times[(p_a^+ )^2m+(p_b^+)^2l]-2[p_a^+ p_b^+  - 2p_a^-p_b^-]\\\nonumber&\times[(p_a^+ )^2m+(p_b^+ )^2l],\\\nonumber
		\kappa_{43}&= 
		(p_a^+)^4 m^2+(p_b^+)^4 l^2
	\end{align}
	with the corresponding characteristic function
	\begin{align}
		R_s(k) = 1 +\frac{1}{4} \int\limits_{-1}^{1}d \lambda_0 \int\limits_{1}^{\infty}d\lambda_1 \int\limits_{1}^{\infty}d\lambda_2 J(\lambda_0,\lambda_1 , \lambda_2) \mathcal{F}_O(\lambda_0, \lambda_1,\lambda_2)(\kappa_1 k +\kappa_2 k^2 +\kappa_3 k^3 +\kappa_4 k^4).
	\end{align}
	
		\subsection*{B --- Channel--factor expansion}
	
	Here, we derive the general channel--factor expansion. For $\beta=1$, we have 
	\begin{align}
		\mathcal{F}_0(\lambda_0,\lambda_1,\lambda_2)&=\prod\limits_{c=1}^{M}\frac{g_c^++\lambda_0}{(g_c^++\lambda_1)^{1/2}(g_c^++\lambda_2)^{1/2}}\\&=\exp \ln \prod\limits_{c=1}^{M}\frac{g_c^++\lambda_0}{(g_c^++\lambda_1)^{1/2}(g_c^++\lambda_2)^{1/2}}\\&=\exp\ln \prod\limits_{c=1}^{M}\frac{1+\frac{T_c}{2}(\lambda_0-1)}{(1+\frac{T_c}{2}(\lambda_1-1))^{1/2}(1+\frac{T_c}{2}(\lambda_2-1))^{1/2}}\\&=
		\exp\biggl(\sum\limits_{c=1}^{M}\sum\limits_{j=1}^{\infty}\frac{(-1)^{j+1}}{j}\biggl(\left(\frac{T_c}{2}(\lambda_0-1)\right)^j-\frac{1}{2}\left(\frac{T_c}{2}(\lambda_1-1)\right)^j-\frac{1}{2}\left(\frac{T_c}{2}(\lambda_2-1)\right)^j\biggr)\biggr)\\&=\exp\left(-\frac{1}{4}(\lambda_1+\lambda_2-2\lambda_0)\sum\limits_{c=1}^{M}T_c\right)	\\&\times\exp\biggl(\sum\limits_{c=1}^{M}\sum\limits_{j=2}^{\infty}\frac{(-1)^{j+1}}{j}\biggl(\left(\frac{T_c}{2}(\lambda_0-1)\right)^j-\frac{1}{2}\left(\frac{T_c}{2}(\lambda_1-1)\right)^j-\frac{1}{2}\left(\frac{T_c}{2}(\lambda_2-1)\right)^j\biggr)\biggr)
	\end{align}
	By using the singularity--free parametrization in the $(p',s',t')$--coordinates, we have 
	\begin{align}
		\mathcal{F}_0(p',s',t')&=\exp\left(-\frac{p'}{4}\sum\limits_{c=1}^{M}T_c\right)	\\&\times\exp\biggl(\sum\limits_{c=1}^{M}\sum\limits_{j=2}^{\infty}\frac{(-1)^{j+1}}{j}\biggl(\left(\frac{T_c}{2}(-p's'^2(1-t'^2))\right)^j\\&-\frac{1}{2}\left(\frac{T_c}{2}(p'(1+s'^2t'^2-2s'^2))\right)^j-\frac{1}{2}\left(\frac{T_c}{2}(p's'^2t'^2)\right)^j\biggr)\biggr).
	\end{align}
	By making use of the exponential series, we see that \begin{align}
		&\exp\biggl(\sum\limits_{c=1}^{M}\sum\limits_{j=2}^{\infty}\frac{(-1)^{j+1}}{j}\biggl(\left(\frac{T_c}{2}(-p's'^2(1-t'^2))\right)^j-\frac{1}{2}\left(\frac{T_c}{2}(p'(1+s'^2t'^2-2s'^2))\right)^j-\frac{1}{2}\left(\frac{T_c}{2}(p's'^2t'^2)\right)^j\biggr)\biggr)\\&= 1+\mathcal{O}(p'^2). 
	\end{align}
	This means that both in the leading and subleading order term of the asymptotic expansion in $1/\Xi$ the term containing nonlinear dependencies, \textit{e.g.} $T_c^j$, where $j\geq 2$, do not contribute. For the derivation of the leading-- and subleading--order term it therefore suffices to approximate the channel factor by 
	\begin{align}
		\mathcal{F}_0(p',s',t')&\simeq\exp\left(-\frac{p'}{4}\sum\limits_{c=1}^{M}T_c\right)=\exp\left(-\frac{p'}{2}\pi\Xi\right).
	\end{align}
	In the limit of infinitely many small channels that scale with $T_c\simeq 1/M$, we get 
	\begin{align}
		&\lim_{M \to \infty} \exp\left(-\frac{p'}{4}\sum\limits_{c=1}^{M}T_c\right)	\exp\biggl(\sum\limits_{c=1}^{M}\sum\limits_{j=2}^{\infty}\frac{(-1)^{j+1}}{j}\biggl(\left(\frac{T_c}{2}(-p's'^2(1-t'^2))\right)^j\\&-\frac{1}{2}\left(\frac{T_c}{2}(p'(1+s'^2t'^2-2s'^2))\right)^j-\frac{1}{2}\left(\frac{T_c}{2}(p's'^2t'^2)\right)^j\biggr)\biggr)\\&= 	\lim_{M \to \infty} \exp\left(-\frac{p'}{4}(2\pi\Xi)\right)	\exp\biggl(\sum\limits_{c=1}^{M}\sum\limits_{j=2}^{\infty}\frac{(-1)^{j+1}}{j}\biggl(\left(\frac{1}{2M}(-p's'^2(1-t'^2))\right)^j\\&-\frac{1}{2}\left(\frac{1}{2M}(p'(1+s'^2t'^2-2s'^2))\right)^j-\frac{1}{2}\left(\frac{1}{2M}(p's'^2t'^2)\right)^j\biggr)\biggr)\\&=\exp\left(-\frac{p'\pi\Xi}{2}\right).
	\end{align}
	The last equation holds since the sum over the transmission coefficients scales linearly with $M$, whereas the term $\left(1/(2M)(p'(1+s'^2t'^2-2s'^2))\right)^j$ at least scales with $1/M^2$ and therefore vanishes in the limit considered. Thus, for infinitely many small channels, the channel--factor exactly turns into the kernel of the Laplace transform with parameter $\pi\Xi/2$.
	\subsection*{C --- Double series representation of the $\psi$-integral}
	In \cite{Matrixelemente}, the integral 
	\begin{align}
2\pi \int\limits_{0}^{2\pi}d\psi I_0\left(2\sqrt{(A_1+B_1\exp{(2\psi i )}+C_2 \exp{(-2\psi i)})(A_2+B_2 \exp{(-2\psi i)}+C_1\exp{(2\psi i)})}\right)
	\end{align}
	 was considered. We present a double series expansion of this integral and show how this can be used to resolve the phase dependencies in the asymptotic expansion.
	Using the series representation of the modified Bessel function, 
	\begin{align}
		I_0(x) = \sum\limits_{n=0}^{\infty} \frac{1}{n! \Gamma(n+1)2^{2n}}x^{2n},
	\end{align}
	we can derive a double series expression for the integral
	\begin{align}
		&2\pi \int\limits_{0}^{2\pi}d\psi I_0\left(2\sqrt{(A_1+B_1\exp{(2\psi i )}+C_2 \exp{(-2\psi i)})(A_2+B_2 \exp{(-2\psi i)}+C_1\exp{(2\psi i)})}\right)\\\nonumber
		&=2\pi \sum\limits_{n=0}^{\infty}\int\limits_{0}^{2\pi}d\psi\frac{1}{n!\Gamma{(n+1)}}\left((A_1+B_1\exp{(2\psi i )}+C_2 \exp{(-2\psi i)})(A_2+B_2 \exp{(-2\psi i)}+C_1\exp{(2\psi i)})\right)^n\\\nonumber
		&=2\pi \sum\limits_{n=0}^{\infty}\int\limits_{0}^{2\pi}d\psi\frac{1}{n!\Gamma(n+1)}\Biggl(A_1A_2+B_1B_2+C_1C_2+\exp{(2\psi i)}(B_1A_2+C_1 A_1)\\\nonumber
		&+\exp{(-2\psi i)}(C_2A_2+B_2A_1)+\exp{(4\psi i)}B_1C_1+\exp{(-4\psi i)}C_2B_2\Biggr)^n.
	\end{align}
	Employing the multinomial theorem
	\begin{align}
		\left(\sum\limits_{j=0}^{J}A_j\right)^n = \sum_{\substack{k_1+...+k_J=n\\k_1,...,k_J\geq0}}\binom{n}{k_1,..,k_J}\prod\limits_{l=1}^{J}A_l^{\substack{k_l}},
	\end{align} we get with $J=5$
	\begin{align}
		&2\pi \sum\limits_{n=0}^{\infty}\int\limits_{0}^{2\pi}d\psi\frac{1}{n!\Gamma{(n+1)}}\Biggl(A_1A_2+B_1B_2+C_1C_2+\exp{(2\psi i)}(B_1A_2+C_1 A_1)\\\nonumber
		&+\exp{(-2\psi i)}(C_2A_2+B_2A_1)+\exp{(4\psi i)}B_1C_1+\exp{(-4\psi i)}C_2B_2\Biggr)^n\\\nonumber
		&2\pi \sum\limits_{n=0}^{\infty}\int\limits_{0}^{2\pi}d\psi \frac{1}{n!\Gamma{(n+1)}}\sum_{\substack{k_1+...+k_5=n\\k_1,...,k_5\geq0}}\binom{n}{k_1,...,k_5}
		(A_1A_2+B_1B_2+C_1C_2)^{k_1}\exp{(2\psi i k_2)}(B_1A_2+C_1A_1)^{k_2}\\\nonumber&\times\exp{(-2\psi i k_3)}(C_2A_2+B_2A_1)^{k_3}\exp{(4\psi i k_4)}(B_1C_1)^{k_4}\exp{(-4\psi i k_5)}(C_2B_2)^{k_5}\\\nonumber&=2\pi \sum\limits_{n=0}^{\infty}\int\limits_{0}^{2\pi}d\psi \frac{1}{n!\Gamma{(n+1)}}\sum_{\substack{k_1+...+k_5=n\\k_1,...,k_5\geq0}}\binom{n}{k_1,...,k_5}
		((A_1A_2+B_1B_2+C_1C_2)^{k_1}(B_1A_2+C_1A_1)^{k_2}\\\nonumber&\times(C_2A_2+B_2A_1)^{k_3}(B_1C_1)^{k_4}(C_2B_2)^{k_5})\exp{(2\psi i (k_2-k_3+2(k_4-k_5))}\\\nonumber&=4\pi^2 \sum\limits_{n=0}^{\infty}\sum_{\substack{k_1+...+k_5=n\\k_2-k_3+2(k_4-k_5)=0\\k_1,...,k_5\geq0}}\frac{1}{n!\Gamma{(n+1)}}\binom{n}{k_1,...,k_5}\\\nonumber&\times
		(A_1A_2+B_1B_2+C_1C_2)^{k_1}(B_1A_2+C_1A_1)^{k_2}(C_2A_2+B_2A_1)^{k_3}(B_1C_1)^{k_4}(C_2B_2)^{k_5}.
	\end{align}
	In the last step, 
	\begin{align}
		\int\limits_{0}^{2\pi}d\psi \exp{(2\psi i n) } = \begin{cases} 2\pi, &n=0\\0, &n \in \mathds{Z}\setminus{\{0\}}
		\end{cases}
	\end{align}
	was used. The phase integral only evaluates to nonzero contributions if the exponential is equal to zero. This allows to further evaluate the characteristic function and moments in an asymptotic limit, as one only has to count the functions that do not carry terms that depend on $\psi$. Also, one can cut off the series including the multinomial coefficients at the desired order in $1/\Xi$, which allows for further analytical evaluation of the characteristic function.
	\subsection*{D --- Explicit calculation of the universal Gaussian}
	Here, we calculate the two--dimensional integral in $x,\ y$ arising in the derivation of the universal Gaussian in \cref{eq:Rsk}. We have
		\begin{align}
		&\int\limits_{0}^{1}dy\ k^2(1-y)(-1+x+y-xy)\Biggl[\exp{\left(-\frac{k^2y}{2(g_a^++1)(g_b^++1)(2\pi\Xi)}\right)} \\\nonumber&\times(k^4(y-1)y(xy-1)+2(g_a^++1)(g_b^++1)k^2(y-1)(2\pi\Xi)\\\nonumber&+4 (g_a^++1)^2(g_b^++1)^2(2\pi\Xi)^2)\\\nonumber&+\exp{\left(-\frac{k^2(1-x(y-1))}{2(g_a^++1)(g_b^++1)(2\pi\Xi)}\right)}(k^4(1+x-xy+(1-x(y-1))^2y)-2(g_a^++1)(g_b^++1)k^2x(1-y)(2\pi\Xi)\\&+4(g_a^++1)^2(g_b^++1)^2(2\pi\Xi)^2\Biggr](16(g_a^++1)^3(g_b^++1)^3(1-y)^2(1-xy)^2(2\pi\Xi)^3)^{-1}\\\nonumber&=k^2(1-x)\Biggl(-\exp{\Biggl(-\frac{k^2 y}{2(g_a^++1)(g_b^++1)(2\pi\Xi)}\Biggr)}x^2y(k^2 (y-1) + 2 (g_a^++1) (g_b^++1) (2\pi\Xi))\\\nonumber&+\exp{\Biggl(-\frac{k^2(1+x-xy)}{2(g_a^++1)(g_b^++1(2\pi\Xi))}\Biggr)}(x(y-1)-1)(k^2x(1-y)+2(g_a^++1)(g_b^++1)(2\pi\Xi))\Biggr)\\&\times(8(g_a^++1)^2(g_b^++1)^2x^2(1-xy)(2\pi\Xi)^2)^{-1}\Biggl|_{0}^{1}
		\\\nonumber&=-k^2 \Biggl[2\exp{\left(-\frac{k^2}{2(g_a^++1)(g_b^++1)(2\pi\Xi)}\right)}(x^2+1)(g_a^++1)(g_b^++1)(2\pi\Xi)\\\nonumber&+\exp{\left(-\frac{k^2(x+1)}{2(g_a^++1)(g_b^++1)(2\pi\Xi)}\right)} (x^2-1)(k^2x+2(g_a^++1)(g_b^++1)(2\pi\Xi)\Biggr)\\& \times(8(g_a^++1)^2(g_b^++1)^2x^2(2\pi\Xi)^2)^{-1}.
	\end{align}
	The remaining integral is done similarly and reads
		\begin{align}
		-k^2 \nonumber&\int\limits_{-1}^{1}dx\Biggl[2\exp{\left(-\frac{k^2}{2(g_a^++1)(g_b^++1)(2\pi\Xi)}\right)}(x^2+1)(g_a^++1)(g_b^++1)(2\pi\Xi)\\\nonumber&+\exp{\left(-\frac{k^2(x+1)}{2(g_a^++1)(g_b^++1)(2\pi\Xi)}\right)} (x^2-1)(k^2x+2(g_a^++1)(g_b^++1)(2\pi\Xi))\Biggr]\\& \times(8(g_a^++1)^2(g_b^++1)^2x^2(2\pi\Xi)^2)^{-1}\\\nonumber&=
		\exp{\left(-\frac{k^2(x+1)}{2(g_a^++1)(g_b^++1)(2\pi\Xi)}\right)}\Biggl[1-\exp{\left(-\frac{k^2x}{2(g_a^++1)(g_b^++1)(2\pi\Xi)}\right)}\Biggr]k^2\frac{(x^2-1)}{4(g_a^++1)(g_b^++1)x (2\pi\Xi) }\\\nonumber&+\exp{\left(-\frac{k^2(x+1)}{2(g_a^++1)(g_b^++1)(2\pi\Xi)}\right)}\Biggl|_{-1}^{1} \\&= \exp{\left(-\frac{k^2}{(g_a^++1)(g_b^++1)(2\pi\Xi)}\right)}-1, 
	\end{align}
	which is the result presented in the main text. 
	\subsection*{E --- Cross--section distribution}
Here, we give the integrand from \cref{eq:bichar} as derived in \cite{Wirkungsquerschnitte}.
	\begin{align}
		p_c^j=\frac{|\mathbf{k}|}{8}\frac{\sqrt{|\lambda_j^2-1|}}{(g_c^{+} +\lambda_j)},~~~~ j=0,1,2,
	\end{align}
	\begin{align}
		p_c^{\pm}=p_c^1\pm p_c^2,
	\end{align}
	\begin{align}
		\label{qcp}
		q_c^{+}=\frac{\mathbf{k}}{8i}\left(\frac{E}{\sqrt{4v^2-E^2}}+i g_c^{-}\right)\left(\frac{1}{g_c^{+}+\lambda_1}+\frac{1}{g_c^{+}+\lambda_2}
		-\frac{2}{g_c^{+}+\lambda_0}\right),
	\end{align}
	\begin{align}
		q_c^{-}=\frac{\mathbf{k}}{8i}\left(\frac{E}{\sqrt{4v^2-E^2}}+i g_c^{-}\right)\left(\frac{1}{g_c^{+}+\lambda_1}-\frac{1}{g_c^{+}+\lambda_2}\right),
	\end{align}
	\begin{align}
		r_c^{+}=\frac{i\mathbf{k}^*}{8}\left(\frac{E}{\sqrt{4v^2-E^2}}-i g_c^{-}\right)\left(\frac{1}{g_c^{+}+\lambda_1}+\frac{1}{g_c^{+}+\lambda_2}
		-\frac{2}{g_c^{+}+\lambda_0}\right),
	\end{align}
	\begin{align}
		r_c^{-}=\frac{i\mathbf{k}^*}{8}\left(\frac{E}{\sqrt{4v^2-E^2}}-i g_c^{-}\right)\left(\frac{1}{g_c^{+}+\lambda_1}-\frac{1}{g_c^{+}+\lambda_2}\right).
	\end{align}
	It is  $r_c^{\pm}=(q_c^{\pm})^*$. We further need the quantities $l=X/Y$, $m=Y/X$, $\omega=2\sqrt{X Y}$, where
	\begin{align}
		X=2p_a^{+}+q_a^{-}e^{-i2\psi}+r_a^{-}e^{i2\psi},~~~~
		Y=2p_b^{+}+q_b^{-}e^{i2\psi}+r_b^{-}e^{-i2\psi}.
	\end{align}
 The $\kappa$'s are given as
	\begin{eqnarray}
		\nonumber
		\kappa_1=\kappa_{11} J_1(\omega),
		~~~ 
		\kappa_2=\kappa_{21} J_0( \omega)+\kappa_{22} J_2( \omega),\hspace{1.8cm}\\
		\kappa_3=\kappa_{31} J_1(\omega)+\kappa_{32} J_3(\omega),
		~~~
		\kappa_4=\kappa_{41} J_0(\omega)+\kappa_{42} J_2(\omega)+\kappa_{43} J_4(\omega).
	\end{eqnarray}
	~\\
	The coefficients of the Bessel functions above are as follows
	\begin{align}
		\kappa_{11}=-(9/8)\{p_a^{+}m^{1/2} \}_{+},
	\end{align}
	\begin{eqnarray}
		\kappa_{21}=-(1/4)(128 p_a^0p_b^0+14p_a^{+}p_b^{+}+32p_a^{-}p_b^{-})
		-\{3e^{i2\psi}(p_a^{-}q_b^{+}+p_b^{-}r_a^{+})\}_{+}
		-\{e^{-4i\psi}q_a^{-}r_b^{-}\}_{+},
	\end{eqnarray}
	\begin{align}
		\kappa_{22}=-(1/4)\{(p_a^{+}p_a^{+}-4q_a^{-}r_a^{-})m\}_{+}, 
	\end{align}
	\begin{eqnarray}
		\nonumber
		&&\kappa_{31}=\big\{-2\big[(p_a^{+}p_a^{+}+q_a^{-}r_a^{-})m^{1/2}+2(8p_a^0p_b^0+p_a^{+}p_b^{+}+p_a^{-}p_b^{-})l^{1/2}\big](e^{i2\psi}q_b^{-}+e^{-i2\psi}r_b^{-})\\
		\nonumber
		&&+2\big[(p_a^{+}p_b^{-}+4p_a^{-}p_b^{+})m^{1/2}+p_b^{+}p_b^{-}l^{1/2}\big](e^{-i2\psi}q_a^{+}+e^{i2\psi}r_a^{+})+
		\big[16p_a^0(2p_a^0 p_b^{+}-3p_b^0 p_a^{+})\\
		\nonumber
		&&
		+6p_a^{+}(q_a^{+} q_b^{+}+r_a^{+} r_b^{+})+2p_b^{+}(4q_a^{+} r_a^{+}-q_a^{-} r_a^{-})-4p_a^{-}(p_a^{+}p_b^{-}-2p_a^{-}p_b^{+})-3p_a^{+}(q_a^{-}q_b^{-}+r_a^{-}r_b^{-}+p_a^{+}p_b^{+})
		\\
		\nonumber
		&&-(e^{-i4\psi}/2)q_a^{-}(4p_a^{+}r_b^{-}+3p_b^{+}q_a^{-}+2e^{-i2\psi}q_a^{-}r_b^{-}-8e^{i2\psi}r_a^{+}r_b^{+})\\
		&&-(e^{i4\psi}/2)r_a^{-}(4p_a^{+}q_b^{-}+3p_b^{+}r_a^{-}+2e^{i2\psi}q_b^{-}r_a^{-}-8e^{-i2\psi}q_a^{+}q_b^{+})\big]m^{1/2}\big\}_{+},
	\end{eqnarray}
	\begin{eqnarray}
		\kappa_{32}=\big\{p_a^{+}\big[(p_a^{+}p_a^{+}+2q_a^{-}r_a^{-})
		+(3/2)(e^{-i4\psi} q_a^{-}q_a^{-}+e^{i4\psi}r_a^{-}r_a^{-})
		+(2p_a^{+}p_a^{+}+q_a^{-}r_a^{-})(e^{-i2\psi} q_a^{-}+e^{i2\psi}r_a^{-})\big]m^{3/2}\big\}_{+},
	\end{eqnarray}
	\begin{eqnarray}
		\nonumber
		\kappa_{41}\!\!\!&=&\!\!\!32 \big[2 p_a^0 p_a^0(p_b^{-}+e^{i2\psi}q_b^{+})(p_b^{-}+e^{-i2\psi}r_b^{+})
		+2 p_b^0 p_b^0(p_a^{-}+e^{-i2\psi}q_a^{+})(p_a^{-}+e^{i2\psi}r_a^{+})\\
		\nonumber
		&+&\!\!\! p_a^0 p_b^0\big((p_a^{+}+e^{-i2\psi}q_a^{-})(p_b^{+}+e^{-i2\psi}r_b^{-})+(p_a^{+}+e^{i2\psi}r_a^{-})(p_b^{+}+e^{i2\psi}q_b^{-})\big)\big]\\
		\nonumber
		&+&\!\!\! 256 p_a^0 p_a^0p_b^0 p_b^0
		+(p_a^{+}+e^{-i2\psi}q_a^{-})^2(p_b^{+}+e^{-i2\psi}r_b^{-})^2
		+(p_a^{+}+e^{i2\psi}r_a^{-})^2(p_b^{+}+e^{i2\psi}q_b^{-})^2\\
		\nonumber
		&+&\!\!\! 4\big[(p_a^{+}+e^{-i2\psi}q_a^{-})(p_b^{+}+e^{i2\psi}q_b^{-})
		-2(p_a^{-}+e^{-i2\psi}q_a^{+})(p_b^{-}+e^{i2\psi}q_b^{+})\big]\\
		&&\times\big[(p_a^{+}+e^{i2\psi}r_a^{-})(p_b^{+}+e^{-i2\psi}r_b^{-})
		-2(p_a^{-}+e^{i2\psi}r_a^{+})(p_b^{-}+e^{-i2\psi}r_b^{+})\big],
	\end{eqnarray}
	\begin{eqnarray}
		\nonumber
		\kappa_{42}=-32p_a^0p_b^0\big[(p_a^{+}+e^{-i2\psi}q_a^{-})(p_a^{+}+e^{i2\psi}r_a^{-})m+(p_b^{+}+e^{i2\psi}q_b^{-})(p_b^{+}+e^{-i2\psi}r_b^{-}) l ]\hspace{2cm}\\
		\nonumber
		+2\big[(p_a^{+}+e^{-i2\psi}q_a^{-})(p_b^{+}+e^{i2\psi}q_b^{-})
		-2(p_a^{-}+e^{-i2\psi}q_a^{+})(p_b^{-}+e^{i2\psi}q_b^{+})\big]
		\big[(p_a^{+}+e^{i2\psi}r_a^{-})^2 m
		+(p_b^{+}+e^{-i2\psi}r_b^{-})^2 l\big]\\
		+ 2\big[(p_a^{+}+e^{i2\psi}r_a^{-})(p_b^{+}+e^{-i2\psi}r_b^{-})
		-2(p_a^{-}+e^{i2\psi}r_a^{+})(p_b^{-}+e^{-i2\psi}r_b^{+})\big]
		\big[(p_a^{+}+e^{-i2\psi}q_a^{-})^2 m
		+(p_b^{+}+e^{i2\psi}q_b^{-})^2 l \big],\ 
	\end{eqnarray}
	\begin{eqnarray}
		\kappa_{43}=\big\{(p_a^{+}+e^{-i2\psi}q_a^{-})^2(p_a^{+}+e^{i2\psi}r_a^{-})^2 m^2\big\}_{+}.
	\end{eqnarray}
	In the above equations,  an expression $\mathcal{E}$ including  $a,b,l,m,\psi$ in $\{\,\}_{\pm}$ is given by $\{\mathcal{E}(a,b,l,m,\psi)\}_{\pm}
	:=\mathcal{E}(a,b,l,m,\psi)\pm\mathcal{E}(b,a,m,l,-\psi)$. To calculate the asymptotic expansion, we apply the coordinate transformation from Eqs. (7)--(9) and expand all prefactors of the Bessel functions in a Taylor series around $p'=0$. 
	\newpage
	\bibliography{EricsonGOE}
	\end{document}